\documentclass[aps,pre,twocolumn,superscriptaddress,floatfix]{revtex4-2}
\usepackage{amsmath}
\usepackage{amsfonts}
\usepackage{amssymb}
\usepackage{lmodern,dsfont}
\usepackage{graphicx}
\usepackage[usenames,dvipsnames]{xcolor}
\usepackage{bm}
\usepackage[english=american]{csquotes}

\newcommand{\Av}[1]{\left\langle #1 \right\rangle}
\newcommand{\av}[1]{\langle #1 \rangle}
\newcommand{\n}{\nonumber}
\newcommand{\nn}{\nonumber \\}

\newcommand{\grad}{\bm{\nabla}}
\renewcommand{\eqref}[1]{Eq.~(\ref{#1})}

\begin{document}

\author{Andreas Dechant}
\affiliation{Department of Physics \#1, Graduate School of Science, Kyoto University, Kyoto 606-8502, Japan}
\title{Bounds on the precision of currents in underdamped Langevin dynamics}
\date{\today}

\begin{abstract}
We derive bounds on the precision of fluctuating currents, which are valid for the steady state of underdamped Langevin dynamics.
These bounds provide a generalization of the overdamped thermodynamic uncertainty relation to the finite-mass regime.
In the overdamped case, the precision of a current is bounded by the entropy production.
By contrast, the underdamped bound acquires two additional positive terms, which characterize the local mean acceleration and the fluctuations of the velocity.
We generalize the bound to the cases of a magnetic field and anisotropic temperature, and derive a joint bound for several observables.
Finally, we apply our results to biased free diffusion and the Brownian gyrator (with and without magnetic field), as well as to diffusion in periodic potentials.
In the latter case, we show that the underdamped bound can be tight when taking into account the correlations between the current and other observables.
\end{abstract}

\maketitle

\section{Introduction}

Currents are the measurable consequence of a physical system being out of equilibrium.
In thermal equilibrium, the principle of detailed balance demands that every possible transition in the system is just as likely as the reverse transition, and thus, the statistical average of any current vanishes.
In out-of-equilibrium systems, by contrast, detailed balance is broken and forward and reverse transitions do not occur with the same probability, which manifests itself in non-vanishing average currents.
The degree to which detailed balance is broken is characterized by the entropy production rate, which is positive and vanishes only in equilibrium.
Thus, measuring a finite average current implies that the entropy production rate is strictly positive.

On the other hand, the probabilistic nature of the transitions implies that, both in and out of equilibrium, the observed value of a current fluctuates.
For macroscopic systems, the magnitude of these fluctuations can generally be neglected, however, for small systems, the fluctuations are an essential contribution to the measured current.
A remarkable result, which has recently received much theoretical and experimental attention, is the thermodynamic uncertainty relation (TUR) \cite{Bar15,Gin16,Pie17,Hor17,Hor20}, originally conjectured by Barato and Seifert \cite{Bar15} and later proven by Gingrich et al.~\cite{Gin16}.
This relation is an inequality that states that the precision of any stochastic current---the ratio of its average squared divided by its variance---is bounded from above by half the amount of entropy production.
This implies that, indeed, the average and the fluctuations of currents are not independent, but their relative size is controlled by the degree to which detailed balance is broken.

The TUR in its original form applies to steady state currents in irreversible continuous-time Markov processes, specifically, Markov jump \cite{Bar15,Gin16,Pie17,Hor17} and overdamped Langevin dynamics \cite{Dec17,Pie18}.
It has subsequently been generalized to systems with time-dependent driving \cite{Bar18,Koy18,Koy19,Koy20,Liu20}, discrete-time Markov processes \cite{Pro17,Has19,Dec20}, quantum-mechanical systems \cite{Has20,Has21} and some classes of non-Markovian dynamics \cite{Ter20}.
One particular class of dynamics, for which, despite recent efforts \cite{Van19,Fis20,Lee19,Lee21,Kwo22}, a straightforward generalization is still lacking are underdamped Langevin dynamics.
The underdamped Langevin equation describes the motion of finite-mass particles in contact with a thermal environment at low or moderate damping.
Since overdamped dynamics arise as the small-mass limit of underdamped dynamics, it is not unreasonable to expect that the latter should also be covered by the TUR, as has been conjectured in Ref.~\cite{Fis20}.
Very recently, however, Pietzonka \cite{Pie21} showed that, by constructing a potential landscape that exploits the coherent oscillations of a finite-mass particle, the TUR can indeed be violated in underdamped systems.

Given these findings, it is natural to ask whether any bounds on the precision of a stochastic current exist for underdamped dynamics and how they are related to entropy production.
One general answer to this question is given in Refs.~\cite{Van19,Lee19}, following the earlier result for one-dimensional periodic systems in Ref.~\cite{Fis18}.
There, it was shown that an upper bound on the current indeed exists.
However, this bound involves a quantity known as dynamical activity \cite{Mae17}, which diverges in the overdamped limit and does not vanish at equilibrium, and thus the bound is not tight for small mass or near equilibrium.
An alternative approach was pursued in Refs.~\cite{Lee21,Kwo22}, where a bound involving only the entropy production was derived.
However, this bound is not an upper bound on the precision of the current, but rather on a more general quantity, which involves the response of the current to a perturbation, which may not always be realizable in physical systems.

In this work, we derive a bound on the precision of a current, which converges to the overdamped TUR in the small-mass limit and vanishes near equilibrium.
In contrast to the overdamped case, where the precision of the current is bounded by the entropy production, we find that, in the underdamped case, the bound is given by the sum of the entropy production and two additional positive terms, which characterize the local mean acceleration and the velocity fluctuations in the system, respectively.
Since the existence of acceleration and non-thermal velocity fluctuations are precisely the features that differentiate the finite-mass from the overdamped regime, it is reasonable that they appear in the bound.
While the local-mean-acceleration term has an explicit expression, the velocity-fluctuation term is expressed in terms of a velocity-potential, which generally has to be determined from the precise form of the velocity statistics in a given system.
However, the second term can also be written explicitly under suitable assumptions, for example when only the first cumulant of the velocity statistics depends on the position.
Importantly, in the small-mass limit, the velocity fluctuation term is explicitly of order $m$, while the local-mean-acceleration term is of order $m^2$, so that both terms vanish in the overdamped limit, recovering the TUR.
Moreover, by investigating several example systems, we find that our bound can be tight for finite mass, which indicates that it is indeed the appropriate generalization of the overdamped TUR.

The paper is organized as follows: In Sec.~\ref{sec-underdamped}, we start with a brief review of underdamped Langevin dynamics and define the local mean velocity and the entropy production.
In Sec.~\ref{sec-tur}, we review the TUR for the overdamped case and recent efforts to extend it to underdamped dynamics.
We then proceed in Sec.~\ref{sec-bounds} to derive two new bounds on the precision of a current for underdamped dynamics.
We discuss how these bounds provide an extension of the TUR to underdamped dynamics and how they converge to the overdamped TUR in the overdamped limit.
In Sec.~\ref{sec-magnetic}, we extend the bounds to the case of magnetic fields.
While it is known that the TUR can be violated in the presence of a magnetic field \cite{Chu19}, we show that a general upper bound on the precision can still be obtained.
In Sec.~\ref{sec-aniso}, we extend the bounds to the case of coordinate-dependent and anisotropic temperature and friction.
Another extension is discussed in Sec.~\ref{sec-multi-tur}, where we develop a multidimensional TUR \cite{Dec18c} and correlation TUR \cite{Dec21} for the underdamped case.
The purpose of Sec.~\ref{sec-finite-perturbation}, rather than being directly concerned with bounds on the precision, is to motivate the derivation of our results from a more general point of view.
Finally, in Sec.~\ref{sec-examples}, we discuss several explicit examples and how our new inequalities apply to them.

\section{Underdamped Langevin dynamics} \label{sec-underdamped}

In this work, we focus on the Langevin equations \cite{Cof17} describing the motion of a particle with position $\bm{x}(t) = (x_1(t),x_2(t),x_3(t))$ and velocity $\bm{v}(t) = (v_1(t),v_2(t),v_3(t))$ diffusing in a viscous environment,
\begin{subequations}
\begin{align}
\dot{\bm{x}}(t) &= \bm{v}(t) \\
m \dot{\bm{v}}(t) &= \bm{F}(\bm{x}(t)) - \gamma \bm{v}(t) + \sqrt{\frac{2 \gamma T}{m}} \bm{\xi}(t) .
\end{align} \label{langevin}%
\end{subequations}
Here, $m$ is the mass of the particle, $\bm{F}(\bm{x})$ is an external force, $\gamma$ is the friction coefficient and $T$ is the temperature of the environment.
The vector $\bm{\xi}(t)$ is composed of independent Gaussian white noises, $\av{\bm{\xi}(t)} = 0$ and $\av{\xi_i(t) \xi_j(s)} = \delta_{i j} \delta(t-s)$.
Equivalently, the dynamics can be described by the Kramers-Fokker-Planck equation for the probability density $p(\bm{x},\bm{v},t)$ \cite{Ris86,Cof17},
\begin{align}
\partial_t p(\bm{x},\bm{v},t) &= \mathcal{L}(\bm{x},\bm{v}) p(\bm{x},\bm{v},t) \quad \text{with} \quad \label{kfp}\\
\mathcal{L}(\bm{x},\bm{v}) &= -\bm{v} \cdot \grad_x - \frac{1}{m} \grad_v \cdot \Big( \bm{F}(\bm{x}) - \gamma \bm{v} - \frac{\gamma T}{m} \grad_v \Big) \n .
\end{align}
We remark that, provided the particles have the same mass and friction coefficient, the same equations can also describe the diffusion of $N$ interacting particles; in this case the force $\bm{F}(\bm{x})$ (now a $3 N$-vector) also includes the interactions between the particles.
In the following, we assume that \eqref{kfp} admits a steady-state solution $\mathcal{L}(\bm{x},\bm{v}) p(\bm{x},\bm{v}) = 0$.
This is generally the case if the force includes a confining external potential force or under spatially periodic boundary conditions.
Here and in the following, we take quantities whose arguments do not include the time $t$ to refer to the steady state of the system.
If the force is given by the gradient of a potential, $\bm{F}(\bm{x}) = - \grad_x U(\bm{x})$, then this steady state satisfies detailed balance and is given by the Boltzmann-Gibbs equilibrium density \cite{Ris86}
\begin{align}
p^\text{eq}(\bm{x},\bm{v}) = \frac{1}{Z^\text{eq}} \exp\bigg[ - \frac{m \Vert \bm{v} \Vert^2}{2T} + \frac{U(\bm{x})}{T} \bigg] \label{equilibrium-density},
\end{align}
where $Z^\text{eq}$ is the equilibrium partition function
\begin{align}
Z^\text{eq} = \bigg(\sqrt{\frac{2 \pi T}{m}} \bigg)^{3N} \int d\bm{x} \ \exp\bigg[-\frac{U(\bm{x})}{T} \bigg].
\end{align}
Here the integral over $\bm{x}$ is taken to be over the entire domain of the system, i.~e.~$\mathbb{R}^{3N}$ in the case of a confining potential and the unit cell in the case of periodic boundary conditions.
In the case of a gradient force, it is clear from \eqref{equilibrium-density} that the local mean velocity
\begin{align}
\bm{\nu}(\bm{x},t) &= \frac{\int d\bm{v} \ \bm{v} \ p(\bm{x},\bm{v},t)}{p_x(\bm{x},t)} \quad \text{with} \\
p_x(\bm{x},t) &= \int d\bm{v} \ p(\bm{x},\bm{v},t) \n ,
\end{align}
vanishes in the equilibrium state, $\bm{\nu}^\text{eq}(\bm{x}) = 0$.
This means that at any position $\bm{x}$, a particle is equally likely to move in any direction, or, in other words, there is no flow at any point in space.
This is a direct consequence of detailed balance.

In general, however, the force $\bm{F}(\bm{x})$ is non-conservative and cannot be written as the gradient of a potential.
Examples include a bias force in a periodic lattice or a torque force applied to particles in a confining potential.
Such forces break detailed balance and drive the system out of equilibrium.
In this case, the steady state probability density $p(\bm{x},\bm{v})$ cannot be written in the form \eqref{equilibrium-density} and, in particular, the local mean velocity $\bm{\nu}(\bm{x})$ generally does not vanish in the steady state.
This means that, even though the system reaches a steady state, this steady state supports local flows at some (or all) points in space.
These flows continuously dissipate energy, leading to a positive steady-state entropy production rate \cite{Spi12},
\begin{align}
\sigma &= \frac{1}{\gamma T} \Av{ \bigg\Vert \gamma \bm{v} + \frac{\gamma T}{m} \grad_v \ln p_v \bigg\Vert^2} \\
&= \frac{\gamma}{m} \bigg( \frac{m \Av{\Vert \bm{v} \Vert^2}}{T} - 3 N \bigg) \n .
\end{align}
We remark that the local mean velocity $\bm{\nu}(\bm{x})$ is the average of the velocity $\bm{v}$ conditioned on the position $\bm{x}$,
\begin{align}
\bm{\nu}(\bm{x}) = \int d\bm{v} \ \bm{v} \ p_v(\bm{v} \vert \bm{x}) \label{meanvel},
\end{align}
where $p_v(\bm{v} \vert \bm{x})$ is the conditional velocity probability density.
Writing $\bm{v} = \bm{\nu}(\bm{x}) + \bm{v} - \bm{\nu}(\bm{x})$, the entropy production rate can be expressed as
\begin{align}
\sigma &= \frac{\gamma }{T} \Av{\big\Vert \bm{\nu} \big\Vert^2} + \frac{1}{\gamma T} \Av{ \bigg\Vert \gamma \big(\bm{v} - \bm{\nu} \big) + \frac{\gamma T}{m} \grad_v \ln p_v \bigg\Vert^2} \nn
&= \frac{\gamma}{T} \Av{\Vert \bm{\nu} \big\Vert^2} + \frac{\gamma}{m} \bigg( \frac{m \Av{\Vert \bm{v} - \bm{\nu} \Vert^2}}{T} - 3 N \bigg) \label{entropy-splitting}.
\end{align}
Both terms in this expression are positive. 
The first term directly measures the magnitude of the local flows in the system.
On the other hand, the second term quantifies how much the fluctuations of the velocity around its local mean value differ from thermal fluctuations described by \eqref{equilibrium-density}.
Either of the two terms may vanish independently, i.~e., both non-zero flows with thermal velocity fluctuations and vanishing flows with non-thermal velocity fluctuations are generally possible.
We further remark on the behavior of the two terms in \eqref{entropy-splitting} in the overdamped limit of vanishing mass $m \rightarrow 0$.
In this limit, the first term converges to the overdamped expression for the entropy production rate.
We thus define the overdamped entropy production as the first term in \eqref{entropy-splitting},
\begin{align}
\sigma^\text{od} = \frac{\gamma}{T} \Av{\Vert \bm{\nu} \big\Vert^2} \label{entropy-over}.
\end{align}
By contrast, the second term vanishes, since the velocity density becomes thermal in the overdamped limit.

\section{Thermodynamic uncertainty relation} \label{sec-tur}
While the entropy production \eqref{entropy-splitting} provides a measure of the detailed balance violation in the system, evaluating it explicitly generally requires knowledge about both the local flows and the local velocity fluctuations.
However, in many cases, this information is not directly accessible in a measurement, but we can only measure certain observables, for example the displacement of the particle, or the work performed against an external load force.
Such observables often take the form of time-integrated stochastic currents, whose general form is
\begin{align}
J(t) = \int_0^t ds \ \bm{w}(\bm{x}(s)) \cdot \bm{v}(s) \label{current},
\end{align}
where $\bm{w}(\bm{x})$ is a weighting function.
For $\bm{w}(\bm{x}) = \hat{\bm{e}}$ with some unit vector $\hat{\bm{e}}$ this yields the displacement of the particle in direction $\hat{\bm{e}}$, for $\bm{w}(\bm{x}) = -\bm{F}^\text{load}$ the work against the load force $\bm{F}^\text{load}$.
In the steady state, the average current is given by
\begin{align}
\av{J(t)} = t \int d\bm{x} \ \bm{\rho}(\bm{x}) \cdot \bm{\nu}(\bm{x}) p_x(\bm{x}) \label{current-average} .
\end{align}
This is precisely the analog of time-integrated currents in overdamped Langevin systems
\begin{align}
J^\text{od}(t) = \int_0^t ds \ \bm{w}(\bm{x}(s)) \circ \dot{\bm{x}}(s),
\end{align}
with the Stratonovich product $\circ$, whose average is likewise given by
\begin{align}
\av{J^\text{od}(t)} = t \int d\bm{x} \ \bm{w}(\bm{x}) \cdot \bm{\nu}(\bm{x}) p_x(\bm{x}) ,
\end{align}
where $\bm{\nu}(\bm{x})$ and $p_x(\bm{x})$ are the corresponding local mean velocity and position probability density of the overdamped system.
The expression \eqref{current-average} shows that a non-vanishing average current requires a non-zero local mean velocity and thus, via \eqref{entropy-splitting}, a finite rate of entropy production.
This reflects the intuitive notion that currents only flow in out-of-equilibrium systems, i.~e.~if detailed balance is broken.

The central question that we want to address in this work is whether the relation between currents and entropy production can be cast in a more quantitative form.
For overdamped Langevin systems in the steady state, the answer is affirmative and is provided by the so-called thermodynamic uncertainty relation (TUR) \cite{Bar15,Gin16,Dec17,Pie17,Hor20}
\begin{align}
\eta^\text{od}_J \equiv \frac{\av{J^\text{od}(t)}^2}{\text{Var}(J^\text{od}(t))} \leq \frac{1}{2} \Delta S^\text{od} \label{TUR}.
\end{align}
Here $\text{Var}(J^\text{od}(t)) = \av{(J^\text{od}(t))^2} - \av{J^\text{od}(t)}^2$ denotes the variance of the current and $\Delta S^\text{od} = \sigma^\text{od} t$ is the total entropy production up to time $t$.
Thus, in the overdamped case, the presence of a non-zero current necessarily requires an amount of dissipation that is greater than twice the ratio of the average of the current squared and its variance.
The TUR has two important implications:
On the one hand, it implies a bound on the precision $\eta^\text{od}_J$ of a current, i.~e.~how large the average current can be compared to its fluctuations \cite{Bar15,Gin16,Dec17,Pie17,Pie18}.
On the other hand, it provides a way to estimate the entropy production using a measurement of the current $r(t)$ \cite{Li19,Man20,Ots20,Van20}.

While the TUR has been proven for overdamped Langevin and Markov jump dynamics, its validity for underdamped Langevin dynamics had been conjectured for periodic systems in Ref.~\cite{Fis20}.
Precisely speaking, the TUR can be violated in underdamped Langevin dynamics at short times \cite{Van19,Fis20}, requiring an additional transient correction on the right-hand side of \eqref{TUR}. 
However, more recently, it has been shown \cite{Pie21} that it is indeed possible to violate the TUR by exploiting the coherent oscillating motion of the particle.
We note that bounds on the precision of a current of the type
\begin{align}
\eta_J \equiv \frac{\av{J(t)}^2}{\text{Var}(J(t))} \leq \frac{1}{2} \Sigma, \label{TUR-general}
\end{align}
in which the right-hand side is replaced by a quantity $\Sigma$ involving the so-called dynamical activity, have been derived also for underdamped Langevin dynamics \cite{Fis18,Van19,Lee19}.
We will discuss some examples in the next section.
However, these bounds have two disadvantages: 
First, they do not reproduce the overdamped TUR \eqref{TUR} in the massless limit; instead the corresponding limit of $\Sigma$ is larger than the overdamped entropy production, so the bound is less tight.
And, second, $\Sigma$ does not vanish in equilibrium, and thus, the bound becomes trivial close to equilibrium.
In the following, our goal is to derive bounds of the type \eqref{TUR-general}, which address these shortcomings, that is, where the quantity $\Sigma$ converges to the entropy production in the overdamped limit and vanishes at equilibrium.

\section{Bounds on currents from the fluctuation-response inequality} \label{sec-bounds}
In this section, we wish to derive generalizations to \eqref{TUR} which are valid for the underdamped dynamics \eqref{langevin}.
To do so, we rely on the virtual perturbation technique of Ref.~\cite{Dec20}.
This approach was also applied to underdamped dynamics in Ref.~\cite{Van19}, resulting in a bound of the type \eqref{TUR-general}, with the quantity $\Sigma$ on the right-hand side given by
\begin{align}
\Sigma &= t \big( 9 \sigma + 4 \Upsilon \big) + \Omega_v \qquad \text{with} \label{bound-velocity} \\
\Upsilon &= \frac{1}{\gamma T} \Av{\Vert \bm{F} \Vert^2} - 3 \frac{\gamma}{T} \Av{\Vert \bm{v} \Vert^2} + 4 \frac{\gamma}{m}, \nn
\Omega_v &= 2 \Av{ \big(\bm{v} \cdot \grad_v \ln p_v(\bm{v} \vert \bm{x}) \big)^2} - 2 N^2 \n .
\end{align}
The term $\Omega_v$, which is not extensive in time, is interpreted as a boundary term, while $\Upsilon$ is related to the dynamical activity.
The term $\Upsilon$ is positive and vanishes neither in the overdamped limit nor for equilibrium systems.
This, together with the factor $9$ appearing in front of the entropy production rate, makes it obvious that \eqref{bound-velocity} does not reduce to \eqref{TUR} in the overdamped limit $m \rightarrow 0$.
The crucial difference between the former result and the present work is that, while in Ref.~\cite{Van19}, the virtual perturbation was chosen in such a way as to rescale the velocity $\bm{v} \rightarrow (1+\epsilon) \bm{v}$, here, we only rescale the local mean velocity $\bm{\nu}(\bm{x}) \rightarrow (1+\epsilon) \bm{\nu}(\bm{x})$, see Sec.~\ref{sec-bounds-perturbation} for the details.
This operation allows us to obtain a quantity on the right-hand side of \eqref{TUR-general}, which provides a bound, which is generally tighter and converges to the entropy production in the overdamped limit.
The first main result of this section is the bound
\begin{align}
&\eta_J \leq  \frac{1}{2} \Sigma \qquad \text{with} \label{main-bound-1} \\
\Sigma &= t \big( \sigma^\text{od} + \Psi  \big) + \Omega_\nu, \nn
\Psi &= \frac{4 m^2}{\gamma T} \Av{\big\Vert \bm{\alpha} \big\Vert^2} + \frac{m^2}{\gamma T} \Av{\big\Vert \grad_v \psi \big\Vert^2}, \nn
\Omega_\nu &= 2 \Av{ \big(\bm{\nu} \cdot \grad_v \ln p_v(\bm{v} \vert \bm{x}) \big)^2} \n .
\end{align}
Here, the vector field $\bm{\alpha}(\bm{x})$ is the local mean acceleration, while the potential $\psi(\bm{x},\bm{v})$ is related to the velocity fluctuations around the local mean.
Compared to the overdamped TUR \eqref{TUR}, \eqref{main-bound-1} contains two additional, time-extensive and positive, terms on the right-hand side, which characterize the local mean acceleration and the fluctuations of the velocity, respectively.
Both terms depend on the spatial derivatives of local mean velocity, which implies that, if a violation of the overdamped TUR occurs, it necessarily has to involve a system which exhibits spatially non-uniform flows.
In addition, there is a transient term $\Omega_\nu$, which accounts for the ballistic scaling of the variance of the current for short times.
To better understand the physical meaning of the additional terms, let us consider the deterministic equation of motion
\begin{align}
\dot{\bm{x}}(t) &= \bm{\nu}(\bm{x}(t)) \label{deterministic-evolution} .
\end{align} 
This describes a particle, whose position coordinate $\bm{x}(t)$ is advected by the local mean velocity field $\bm{\nu}(\bm{x})$.
Differentiating with respect to time, we obtain the corresponding acceleration,
\begin{align}
\ddot{\bm{x}}(t) &= \bm{J}_\nu(\bm{x}(t)) \bm{\nu}(\bm{x}(t)) \equiv \bm{\alpha}(\bm{x}(t)) \label{local-mean-acceleration},
\end{align}
where we introduced the Jacobian matrix of the local mean velocity,
\begin{align}
\big(\bm{J}_\nu(\bm{x})\big)_{ij} = \partial_{x_j} \nu_i(\bm{x}) \label{jacobian}.
\end{align}
Thus, $\bm{\alpha}(\bm{x})$ is the acceleration corresponding to the velocity field $\bm{\nu}(\bm{x})$, and can be interpreted as a local mean acceleration.
We see that $\bm{\alpha}(\bm{x})$ scales quadratically with the magnitude of the local mean velocity and explicitly depends on its spatial derivatives.
The velocity-fluctuation potential $\psi(\bm{x},\bm{v})$, on the other hand, obeys the equation
\begin{align}
\grad_u \cdot \bigg( \Big( \big[ \grad_u \psi(\bm{x},\bm{u} + \bm{\nu}(\bm{x})) \big] &- \bm{J}_\nu(\bm{x}) \bm{u} \Big) q_v(\bm{u} \vert \bm{x}) \bigg) \label{psi-equation-main} \\
& = - \bm{\nu}(\bm{x}) \cdot \grad_x q_v(\bm{u} \vert \bm{x}) \n ,
\end{align}
where $p_v(\bm{v} \vert \bm{x}) = q_v(\bm{v} - \bm{\nu}(\bm{x}) \vert \bm{x})$ and we use the convention that derivatives inside brackets only act on terms enclosed in the same brackets.
From the structure of this equation, it becomes clear that $\psi(\bm{x},\bm{v})$ depends on the fluctuations $\bm{u} = \bm{v} - \bm{\nu}(\bm{x})$ around the local mean velocity $\bm{\nu}(\bm{x})$, hence it characterizes the velocity fluctuations.

The downside of \eqref{main-bound-1} is that the velocity-fluctuation potential $\psi(\bm{x},\bm{v})$ generally does not have an explicit expression, but has to be determined from \eqref{psi-equation-main} for a given solution of \eqref{kfp}.
However, we can obtain an explicit form of the bound under suitable assumptions.
First, for the class of local equilibrium dynamics, whose velocity fluctuations around the local mean are thermal,
\begin{align}
p_v(\bm{v} \vert \bm{x}) = \bigg(\sqrt{\frac{m}{2 \pi T}} \bigg)^{3N} \exp\bigg[ - \frac{m \Vert \bm{v} - \bm{\nu}(\bm{x}) \Vert^2}{2 T} \bigg] ,
\end{align}
we obtain $\psi(\bm{x},\bm{v}) = \bm{u} \cdot \bm{J}_\nu^S(\bm{x}) \bm{u} /2$ and the bound
\begin{subequations}
\begin{align}
\Psi &= \frac{m }{\gamma} \Av{ \text{tr}\Big(\big(\bm{J}_\nu^S\big)^2\Big)} + \frac{4 m^2}{\gamma T} \Av{ \big\Vert \bm{\alpha} \big\Vert^2},  \\
\Omega_\nu &= \frac{2 m}{\gamma} \sigma^\text{od},
\end{align}\label{main-bound-1-local}%
\end{subequations}
where $\bm{J}_\nu^S(\bm{x}) = (\bm{J}_\nu(\bm{x}) + \bm{J}_\nu(\bm{x})^\text{T})/2$ is the symmetric part of the Jacobian of $\bm{\nu}(\bm{x})$ and tr denotes the trace.
In the overdamped limit, the velocity fluctuations also become thermal to leading order and this expression also provides the leading-order correction for vanishing mass,
\begin{align}
\Psi \simeq \frac{m }{\gamma} \Av{ \text{tr}\Big(\big(\bm{J}_\nu^S\big)^2\Big)} + O(m^2) \label{main-bound-1-over} .
\end{align}
In particular, we see that the additional terms on the right-hand side of \eqref{main-bound-1} vanish as order $m$ in the overdamped limit and thus we recover the overdamped TUR \eqref{TUR}.
This is in contrast to the bound \eqref{bound-velocity}, whose right-hand side diverges in this limit.
Further, comparing \eqref{main-bound-1} to \eqref{bound-velocity}, the factor $9$ in front of the entropy production rate is absent, and the entropy production rate is replaced by only the overdamped contribution \eqref{entropy-over}.
Comparing the two time-extensive terms in \eqref{main-bound-1-local}, we note that caution is advised when taking the overdamped limit:
Since the first term is of order $m$ and the second one of order $m^2$, it may be tempting to neglect the second term for small mass.
However, the second term is also of fourth order in the local mean velocity, compared to second order in the first term and overdamped entropy production rate.
Thus, when the system is far from equilibrium and the local mean velocity is large, this term may dominate even if the mass is small.
Second, whenever the velocity statistics only depend on the position via the local mean velocity $\bm{\nu}(\bm{x})$, we find
\begin{align}
\Psi &= \frac{m }{\gamma} \Av{ \text{tr}\Big(\bm{J}_\nu \bm{\Theta} \bm{J}_\nu^\text{T}\Big)} + \frac{4 m^2}{\gamma T} \Av{ \big\Vert \bm{\alpha} \big\Vert^2},\label{main-bound-1-first-cumulant}
\end{align}
while the boundary term $\Omega_\nu$ has the same form as in \eqref{main-bound-1}.
Here, the matrix $\bm{\Theta}$ characterizes the deviations of the kinetic temperature, defined as the variance of the velocity around the local mean velocity, from the physical temperature $T$.
Compared to \eqref{main-bound-1-local} the main difference is that the velocity statistics are not necessarily thermal and the kinetic temperature may be anisotropic.

Our second main results is the bound
\begin{align}
&\eta_J^\text{eq} \equiv \frac{\av{J(t)}^2}{\text{Var}^\text{eq}(J(t))} \leq  \frac{1}{2} \Sigma^\text{eq} \qquad \text{with} \label{main-bound-2} \\
\Sigma^\text{eq} &= t \big( \sigma^\text{od} + \Psi^\text{eq}  \big) + \Omega_\nu^\text{eq}, \nn
\Psi^\text{eq} &= \frac{m }{\gamma} \Av{ \text{tr}\Big(\big(\bm{J}_\nu^S\big)^2\Big)}, \qquad \Omega_\nu^\text{eq} = \frac{2 m}{\gamma} \sigma^\text{od} \n .
\end{align}
The crucial difference between \eqref{main-bound-1} and \eqref{main-bound-2} is that in the latter, the fluctuations of the current $J(t)$ are not evaluated in the dynamics \eqref{langevin}, but in the equilibrium dynamics
\begin{subequations}
\begin{align}
\dot{\bm{x}}(t) &= \bm{v}(t) \\
m \dot{\bm{v}}(t) &= T \grad_x \ln p_x(\bm{x}(t)) - \gamma \bm{v}(t) + \sqrt{\frac{2 \gamma T}{m}} \bm{\xi}(t) .
\end{align} \label{langevin-eq}%
\end{subequations}
It is straightforward to check that \eqref{langevin} and \eqref{langevin-eq} result in the same steady state position density $p_x(\bm{x})$.
However, the velocity statistics of \eqref{langevin-eq} are independent of $\bm{x}$ and given by the equilibrium Boltzmann-Gibbs density,
\begin{align}
p_v^\text{eq}(\bm{v}) = \bigg(\sqrt{\frac{m}{2 \pi T}} \bigg)^{3N} \exp\bigg[ - \frac{m \Vert \bm{v} \Vert^2}{2 T} \bigg] \label{boltzmann-gibbs} ,
\end{align}
and thus \eqref{langevin-eq} obeys detailed balance.
While the average current \eqref{current-average} vanishes for \eqref{langevin-eq}, the fluctuations $\text{Var}^\text{eq}(J(t))$ of $J(t)$ are non-zero.
The advantage of \eqref{main-bound-2} is that the right-hand side has an explicit expression.
Further, in many situations we expect $\text{Var}(J(t)) \geq \text{Var}^\text{eq}(J(t))$, as driving the system out of equilibrium typically leads to increased fluctuations.
In these situations, \eqref{main-bound-2} implies a similar bound on $\text{Var}(J(t))$.

In the remainder of this section, we first review the virtual perturbation technique and its application to underdamped dynamics in Sec.~\ref{sec-bounds-general}.
In Sec.~\ref{sec-bounds-perturbation}, we derive the virtual perturbation that leads to a rescaling of the local mean velocity and the corresponding bound.
We show how \eqref{main-bound-1} can be made explicit for local equilibrium dynamics in Sec.~\ref{sec-bounds-local} and if only some cumulants of the velocity statistics depend on position in Sec.~\ref{sec-bounds-cumulant}.
Finally, in Sec.~\ref{sec-bounds-equilibrium}, we introduce the equivalent equilibrium dynamics and use them to derive \eqref{main-bound-2}.

\subsection{General framework} \label{sec-bounds-general}

In order to derive a generalization of \eqref{TUR} to underdamped dynamics, we use the virtual perturbation approach that we introduced in Ref.~\cite{Dec20}.
In short, the idea is to consider another, slightly perturbed, Langevin dynamics
\begin{subequations}
\begin{align}
\dot{\bm{x}}(t) &= \bm{v}(t), \\
m \dot{\bm{v}}(t) &= \bm{F}(\bm{x}(t)) + \epsilon \bm{f}(\bm{x}(t),\bm{v}(t)) - \gamma \bm{v}(t) + \sqrt{\frac{2 \gamma T}{m}} \bm{\xi}(t), 
\end{align} \label{langevin-mod}%
\end{subequations}
with an additional force $\epsilon \bm{f}(\bm{x},\bm{v})$ and $\epsilon \ll 1$.
This additional force generally changes both the steady state and the expectation of the current \eqref{current-average}.
We define the change in the latter compared to \eqref{langevin} as
\begin{align}
\delta \av{J(t)} = \av{J(t)}_\epsilon - \av{J(t)}_{\epsilon = 0}.
\end{align}
As discussed in Ref.~\cite{Dec20}, this change is bounded from above by the product of the variance of $J(t)$ in the original dynamics \eqref{langevin} and the relative entropy between the path probability densities corresponding to \eqref{langevin} and \eqref{langevin-mod}.
Specifically, we have the inequality
\begin{align}
\frac{2\big(\delta \av{J(t)}\big)^2}{\text{Var}(J(t))} &\leq \frac{t}{\gamma T} \int d\bm{x} \int d\bm{v} \ \big\Vert \epsilon \bm{f}(\bm{x},\bm{v}) \big\Vert^2 p(\bm{x},\bm{v}) \\
& \quad + 2 \int d\bm{x} \int d\bm{v} \ \frac{\big(\delta p(\bm{x},\bm{v}) \big)^2}{p(\bm{x},\bm{v})} + O(\epsilon^3) \n .
\end{align}
Here $\delta p(\bm{x},\bm{v}) = p_{\epsilon}(\bm{x},\bm{v}) - p(\bm{x},\bm{v})$ is the leading order change in the steady state probability density.
This inequality is valid for any force $\bm{f}(\bm{x},\bm{v})$ which allows a linear response treatment.
In Ref.~\cite{Dec20}, the TUR for the overdamped case was derived by constructing a perturbation force $\bm{f}^\text{od}(\bm{x})$ that leads to a proportional rescaling of the current according to $\delta \av{J^\text{od}(t)} \simeq \av{J^\text{od}(t)}$.
In the overdamped case, the force $\bm{f}^\text{od}(\bm{x})$ is proportional to the local mean velocity and thus the right-hand side of the above bound is proportional to the entropy production rate \eqref{entropy-over}, yielding the TUR.
Based on this approach, our goal is to construct the perturbation force $\bm{f}(\bm{x},\bm{v})$ in such a way that it leads to a proportional change in the current,
\begin{align}
\delta \av{J(t)} = \epsilon \av{J(t)} + O(\epsilon^2) \label{current-change} .
\end{align}
If we can find such a force, then we have the inequality
\begin{align}
\eta_J &\leq \frac{t}{2 \gamma T} \int d\bm{x} \int d\bm{v} \ \big\Vert \bm{f}(\bm{x},\bm{v}) \big\Vert^2 p(\bm{x},\bm{v}) \label{fri-bound} \\
& \qquad + \int d\bm{x} \int d\bm{v} \ \frac{\big(\pi(\bm{x},\bm{v}) \big)^2}{p(\bm{x},\bm{v})} \n ,
\end{align}
where defined the first-order correction to the steady state density $\delta p(\bm{x},\bm{v}) = \epsilon \pi(\bm{x},\bm{v})$.
The next step is to relate the right-hand side of \eqref{fri-bound} to the entropy production rate \eqref{entropy-splitting}.

\subsection{Local-mean velocity rescaling perturbation} \label{sec-bounds-perturbation}
In order to find a force realizing \eqref{current-change}, we write the leading order change in the average current explicitly
\begin{align}
\delta \av{J(t)} \simeq t \epsilon \int d\bm{x} \int d\bm{v} \ \bm{w}(\bm{x}) \cdot \bm{v} \ \pi(\bm{x},\bm{v})  \label{current-change-2}.
\end{align}
One choice for $\pi(\bm{x},\bm{v})$ that realizes \eqref{current-change} is
\begin{align}
\pi(\bm{x},\bm{v}) = -\bm{\nu}(\bm{x}) \cdot \grad_v p(\bm{x},\bm{v}) \label{probability-mod} .
\end{align}
In Sec.~\ref{sec-finite-perturbation}, we show that this choice arises in a natural way when rescaling the local mean velocity.
Plugging this into \eqref{current-change-2} and integrating by parts with respect to $\bm{v}$, we obtain
\begin{align}
\delta \av{J(t)} \simeq t \epsilon \int d\bm{x} \int d\bm{v} \ \bm{w}(\bm{x}) \cdot \bm{\nu}(\bm{x}) \ p(\bm{x},\bm{v}),
\end{align}
which is precisely \eqref{current-change}.
Also, integrating \eqref{probability-mod} with respect to $\bm{v}$, we find
\begin{align}
\delta p_x(\bm{x}) = 0,
\end{align}
so that the position probability density is unchanged to leading order, which we use in Sec.~\ref{sec-multi-tur}.
By expanding the Kramers-Fokker-Planck equation \eqref{kfp} corresponding to \eqref{langevin-mod} with respect to $\epsilon$, we obtain to first order in $\epsilon$
\begin{align}
\mathcal{L}(\bm{x},\bm{v}) \pi(\bm{x},\bm{v}) = \frac{1}{m} \grad_v \Big( \bm{f}(\bm{x},\bm{v}) p(\bm{x},\bm{v}) \Big)  \label{kfp-first-order}.
\end{align}
Plugging in \eqref{probability-mod} and simplifying, we obtain an equation for $\bm{f}(\bm{x},\bm{v})$,
\begin{align}
\grad_v \cdot \Big( &\bm{f}(\bm{x},\bm{v}) p(\bm{x},\bm{v}) \Big) \label{f-equation} \\
&= \grad_v \cdot \Big(\big(  \gamma \bm{\nu}(\bm{x}) + m \bm{J}_\nu(\bm{x}) \bm{v} \big) p(\bm{x},\bm{v}) \Big) \nn
& \qquad  - m \grad_x \cdot \big( \bm{\nu}(\bm{x}) p(\bm{x},\bm{v}) \big) \n .
\end{align}
Writing the force as
\begin{align}
\bm{f}(\bm{x},\bm{v}) = \gamma \bm{\nu}(\bm{x}) + m \bm{h}(\bm{x},\bm{v}), \label{force-h} 
\end{align}
this translates into an equation for $\bm{h}(\bm{x},\bm{v})$,
\begin{align}
\grad_v \cdot \Big( \bm{h}(\bm{x},\bm{v}) p(\bm{x},\bm{v}) \Big) &= \grad_v \cdot \Big( \bm{J}_\nu(\bm{x}) \bm{v} p(\bm{x},\bm{v}) \Big) \nn
& \quad - \grad_x \cdot \Big( \bm{\nu}(\bm{x}) p(\bm{x},\bm{v}) \Big) \label{h-equation}.
\end{align}
While this equation cannot be solved explicitly in full generality, it allows us to write \eqref{fri-bound} as
\begin{align}
2 \eta_J \leq t \sigma^\text{od} + \frac{m^2 t}{\gamma T} \Av{\big\Vert \bm{h} \big\Vert^2} + 2 \Av{\big(\bm{\nu} \cdot \grad_v \ln p \big)^2} \label{generalized-TUR}  .
\end{align}
Here we used that $\bm{h}(\bm{x},\bm{v})$ and $\bm{\nu}(\bm{x})$ are orthogonal in the sense that
\begin{align}
\Av{ \bm{\nu} \cdot \bm{h}} = 0 \label{orthogonality-h},
\end{align}
see below for the proof of this relation.
The right-hand side of \eqref{generalized-TUR} consists of three positive terms.
The first one is precisely the contribution of the local mean velocity to the entropy production \eqref{entropy-splitting}.
The second one contains the unknown vector field $\bm{h}(\bm{x},\bm{v})$, which is determined by \eqref{h-equation}.
Both of these terms are proportional to time, in contrast to the third term, which represents a transient contribution.
In order to obtain \eqref{main-bound-1}, we first note that \eqref{h-equation} only depends on the conditional velocity probability density. 
Integrating \eqref{kfp} over $\bm{x}$, we get the condition for the steady state
\begin{align}
\grad_x \cdot \big( \bm{\nu}(\bm{x}) p_x(\bm{x}) \big) = 0 \label{stationary-condition}.
\end{align}
Using this and writing $p(\bm{x},\bm{v}) = p_v(\bm{v}\vert \bm{x}) p_x(\bm{x})$, we find
\begin{align}
\grad_v \cdot \Big( \bm{h}(\bm{x},\bm{v}) p_v(\bm{v}\vert \bm{x}) \Big) &= \grad_v \cdot \Big( \bm{J}_\nu(\bm{x}) \bm{v}  p_v(\bm{v}\vert \bm{x}) \Big) \nn
& \quad - \bm{\nu}(\bm{x}) \cdot \grad_x p_v(\bm{v}\vert \bm{x}) \label{h-equation-2}.
\end{align}
Since $\bm{\nu}(\bm{x})$ is also determined by $p_v(\bm{v} \vert \bm{x})$ (see \eqref{meanvel}), we conclude that $\bm{h}(\bm{x},\bm{v})$ is determined solely by the local velocity velocity statistics $p_v(\bm{v}\vert \bm{x})$ and independent of the spatial distribution $p_x(\bm{x})$.
While $\bm{h}(\bm{x},\bm{v})$ generally cannot be obtained explicitly, we can calculate its local mean value by multiplying \eqref{h-equation-2} with $\bm{v}$ and integrating over $\bm{v}$,
\begin{align}
\bar{\bm{h}}(\bm{x}) \equiv \int d\bm{v} \ \bm{h}(\bm{x},\bm{v}) p_v(\bm{v} \vert \bm{x}) = 2 \bm{\alpha}(\bm{x}) \label{h-local-mean} .
\end{align}
Thus, the local mean value $\bar{\bm{h}}(\bm{x})$ is precisely twice the local mean acceleration \eqref{local-mean-acceleration}.
We remark that in the steady state, the local mean velocity and acceleration are orthogonal on average.
To see this, we note
\begin{align}
&\frac{1}{2} \grad_x \cdot \Big( \big\Vert \bm{\nu}(\bm{x}) \big\Vert^2 \bm{\nu}(\bm{x}) p_x(\bm{x}) \Big) \\
& = \frac{1}{2} \big\Vert \bm{\nu}(\bm{x}) \big\Vert^2 \grad_x \cdot \big( \bm{\nu}(\bm{x}) p_x(\bm{x}) \big) + \bm{\nu}(\bm{x}) \cdot \bm{J}_\nu(\bm{x}) \bm{\nu}(\bm{x}) p_x(\bm{x}) \n .
\end{align}
The first term vanishes in the steady state due to \eqref{stationary-condition}.
The left-hand is a total derivative, so it vanishes when integrating over $\bm{x}$ for natural or periodic boundary conditions.
Then, we have
\begin{align}
0 = \Av{ \bm{\nu} \cdot \bm{J}_\nu \bm{\nu}} = \Av{\bm{\nu} \cdot \bm{\alpha}} \label{orthogonality} .
\end{align}
Since the local mean velocity depends only on $\bm{x}$, we may replace $\bm{h}(\bm{x},\bm{v})$ in \eqref{orthogonality-h} by its local mean value and thus the relation follows from \eqref{orthogonality}. 
Writing $\bm{h}(\bm{x},\bm{v}) = 2 \bm{\alpha}(\bm{x}) + \tilde{\bm{h}}(\bm{x},\bm{v})$, the second term in \eqref{generalized-TUR} can be written as
\begin{align}
\frac{m^2 t}{\gamma T} \Av{\big\Vert \bm{h} \big\Vert^2} = \frac{ m^2 t}{\gamma T} \Big( 4 \Av{\big\Vert \bm{\alpha} \big\Vert^2} + \Av{\big\Vert \tilde{\bm{h}} \big\Vert^2} \Big).
\end{align}
The vector field $\tilde{\bm{h}}(\bm{x},\bm{v})$ obeys the equation
\begin{align}
&\grad_v \cdot \Big( \tilde{\bm{h}}(\bm{x},\bm{v}) p_v(\bm{v}\vert \bm{x}) \Big) \label{h-equation-3} \\
&= \grad_v \cdot \Big( \bm{J}_\nu(\bm{x}) \big( \bm{v} - 2\bm{\nu}(\bm{x}) \big) p_v(\bm{v}\vert \bm{x}) \Big) - \bm{\nu}(\bm{x}) \cdot \grad_x p_v(\bm{v}\vert \bm{x}) \n .
\end{align}
Next, we introduce the conditional velocity probability density relative to the local mean velocity
\begin{align}
q_v(\bm{u}\vert \bm{x}) = p_v(\bm{u}+\bm{\nu}(\bm{x}) \vert \bm{x}) \label{conditional-density-relative}
\end{align}
with $\bm{u} = \bm{v} - \bm{\nu}(\bm{x})$.
With this definition, we have
\begin{align}
\bm{\nu}(\bm{x}) \cdot \grad_x p_v(\bm{v}\vert \bm{x}) &= \bm{\nu}(\bm{x}) \cdot \grad_x q_v(\bm{u} \vert \bm{x}) \\
&\quad - \bm{J}_\nu(\bm{x}) \bm{\nu}(\bm{x})\cdot \grad_u q_v(\bm{u}\vert \bm{x})  \n .
\end{align}
Plugging this into \eqref{h-equation-3} and expressing the resulting equation in terms of $\bm{u}$, we obtain
\begin{align}
&\grad_u \cdot \Big( \bm{k}(\bm{x},\bm{u}) q_v(\bm{u}\vert \bm{x}) \Big) \label{h-equation-4} \\
&= \grad_u \cdot \Big( \bm{J}_\nu(\bm{x}) \bm{u} q_v(\bm{u}\vert \bm{x}) \Big) - \bm{\nu}(\bm{x}) \cdot \grad_x q_v(\bm{u}\vert \bm{x}) \n ,
\end{align}
where $\bm{k}(\bm{x},\bm{u}) = \tilde{\bm{h}}(\bm{x},\bm{u}+\bm{\nu}(\bm{x}))$.
This equation is underdetermined:
We only have a single scalar equation for the $3 N$ components of the vector field $\bm{k}(\bm{x},\bm{u})$.
As a consequence, the solution of this equation is not unique.
One particular solution is given by
\begin{align}
\bm{k}(\bm{x},\bm{u}) = \bm{J}_\nu(\bm{x}) \bm{u} - \frac{\bm{\nu}(\bm{x}) \cdot \grad_x \bm{Q}_v(\bm{u} \vert \bm{x})}{q_v(\bm{u}\vert \bm{x})},
\end{align}
where the components of $\bm{Q}_v(\bm{u} \vert \bm{x})$ are given by
\begin{align}
Q_{v,i}(\bm{u} \vert \bm{x}) = \int_{-\infty}^{u_i} dw_i \ q_v(\bm{u}^{(i)}\vert \bm{x}),
\end{align}
where $\bm{u}^{(i)}$ is the vector $\bm{u}$ with its $i$-th component replaced by $w_i$.
$Q_{v,i}(\bm{u} \vert \bm{x})$ may be interpreted as the cumulative velocity probability in direction $i$.
While this shows that a solution to \eqref{h-equation-3} can always be constructed, the resulting bound is not particularly tight.
Going back to \eqref{fri-bound}, we can search for the force $\bm{f}(\bm{x},\bm{v})$ that minimizes the right-hand side while satisfying \eqref{f-equation}.
This is a standard optimization problem (see Appendix \ref{app-optimization} for details) and the resulting condition for a stationary point is that $\bm{f}(\bm{x},\bm{v})$ should be the velocity gradient of a scalar function, $\bm{f}(\bm{x},\bm{v}) = \grad_v \psi_f(\bm{x},\bm{v})$.
Since this is true for any vector field that is independent of $\bm{v}$, $\bm{\mu}(\bm{x}) = \grad_v ( \bm{\mu}(\bm{x}) \cdot \bm{v})$, the first term, $\gamma \bm{\nu}(\bm{x})$, in \eqref{force-h}, as well as term involving the local mean acceleration, $2 m \bm{\alpha}(\bm{x})$, already satisfy this condition.
Then, we only have to impose this condition on the vector field $\bm{k}(\bm{x},\bm{u}) = \grad_u \phi(\bm{x},\bm{u})$.
We then have from \eqref{h-equation-4},
\begin{align}
&\grad_u \cdot \Big( \big[\grad_u \phi(\bm{x},\bm{u}) \big] q_v(\bm{u}\vert \bm{x}) \Big) \label{h-equation-5} \\
&= \grad_u \cdot \Big( \bm{J}_\nu(\bm{x}) \bm{u} q_v(\bm{u}\vert \bm{x}) \Big) - \bm{\nu}(\bm{x}) \cdot \grad_x q_v(\bm{u}\vert \bm{x}) \n ,
\end{align}
which is equivalent to \eqref{psi-equation-main}.
Defining $\psi(\bm{x},\bm{v}) = \phi(\bm{x},\bm{v}-\bm{\nu}(\bm{x}))$, we then obtain our main result, \eqref{main-bound-1}.
We note that \eqref{h-equation-5} is a second-order differential equation for the scalar function $\phi(\bm{x},\bm{u})$, and thus its solution is unique given suitable boundary conditions.
Using \eqref{h-equation-5}, we can write the velocity-fluctuation term in \eqref{main-bound-1} more explicitly as
\begin{align}
\Av{\big\Vert \grad_v \psi \big\Vert^2} = \int d\bm{x} \int d\bm{u} \ \big\Vert \grad_u \phi(\bm{x},\bm{u}) \big\Vert^2 q_v(\bm{u} \vert \bm{x}) p_x(\bm{x}) \label{velocity-fluctuation-term} .
\end{align}
This form explicitly shows that only the fluctuations of the velocity around the local mean value $\bm{u} = \bm{v} - \bm{\nu}(\bm{x})$ contribute to this term.
In summary, we have shown that a force
\begin{align}
\bm{f}(\bm{x},\bm{v}) = \gamma \bm{\nu}(\bm{x}) + 2 m \bm{\alpha}(\bm{x}) + m \grad_v \phi(\bm{x},\bm{v}-\bm{\nu}(\bm{x}))
\end{align}
leads to the change \eqref{probability-mod} in the probability density and thus to a rescaling of the local mean velocity, leading to the bound \eqref{main-bound-1}.

\subsection{Local equilibrium dynamics} \label{sec-bounds-local}
While \eqref{h-equation-5} determines the velocity-fluctuation potential $\psi(\bm{x},\bm{v})$ entering \eqref{main-bound-1}, its solution cannot be written in closed form for general velocity statistics.
However, for many examples of explicitly solvable linear systems, see Sec.~\ref{sec-examples}, the conditional velocity probability density can be written in the form
\begin{align}
p_v(\bm{v} \vert \bm{x}) = \bigg(\sqrt{\frac{m}{2 \pi T}} \bigg)^{3N} \exp\bigg[ - \frac{m \Vert \bm{v} - \bm{\nu}(\bm{x}) \Vert^2}{2 T} \bigg] \label{local-equilibrium}.
\end{align}
In this case, the velocity fluctuations are locally thermal, and we therefor refer to these systems as local equilibrium dynamics. 
It is easily seen that the second term in the entropy production \eqref{entropy-splitting} vanishes and we have $\sigma = \sigma^\text{od}$.
Nevertheless, detailed balance is broken due to the non-zero local mean velocity.
In this case, the velocity statistics relative to the local mean velocity are independent of $\bm{x}$ and \eqref{h-equation-5} yields
\begin{align}
&\grad_u \cdot \Big( \big(\big[\grad_u \phi(\bm{x},\bm{u}) \big] - \bm{J}_\nu(\bm{x}) \bm{u} \big) q_v(\bm{u}) \Big) = 0 \label{potential-condition-local-equilibrium} ,
\end{align}
where $q_v(\bm{u})$ is the Boltzmann-Gibbs equilibrium density \eqref{boltzmann-gibbs}.
Expanding the derivative, we obtain
\begin{align}
\grad_u^2 \phi(\bm{x},&\bm{u}) - \text{tr}\big(\bm{J}_\nu(\bm{x})\big) \\
& - \frac{m}{T} \Big( \bm{u} \cdot \grad_u \phi(\bm{x},\bm{u}) - \bm{u} \cdot \bm{J}_\nu(\bm{x}) \bm{u}\Big) = 0 . \n 
\end{align}
This is solved by the quadratic form
\begin{align}
\phi(\bm{x},\bm{u}) = \frac{1}{2} \bm{u} \cdot \bm{J}_\nu(\bm{x}) \bm{u} = \frac{1}{2} \bm{u} \cdot \bm{J}^S_\nu(\bm{x}) \bm{u} \label{potential-solution-local-equilibrium} ,
\end{align}
where only the symmetric part $\bm{J}^S_\nu(\bm{x}) = (\bm{J}_\nu(\bm{x}) + \bm{J}_\nu(\bm{x})^\text{T})/2$ of the Jacobian contributes.
Plugging this into \eqref{velocity-fluctuation-term}, we find
\begin{align}
\Psi = \frac{m}{\gamma} \Av{ \text{tr}\Big(\big(\bm{J}_\nu^S\big)^2\Big)} + \frac{4 m^2}{\gamma T} \Av{ \big\Vert \bm{\alpha} \big\Vert^2} \label{generalized-TUR-local}.
\end{align}
For the boundary term, we use
\begin{align}
\grad_v \ln p_v(\bm{v} \vert \bm{x}) = -\frac{m}{T} \big(\bm{v} - \bm{\nu}(\bm{x}) \big)
\end{align}
to obtain
\begin{align}
\Omega_\nu = \frac{2 m}{T} \Av{\big\Vert \bm{\nu} \big\Vert^2} = \frac{2 m}{\gamma} \sigma^\text{od} .
\end{align}
This is precisely \eqref{main-bound-1-local}.

\subsection{Position-independent velocity cumulants} \label{sec-bounds-cumulant}
While the local equilibrium density \eqref{local-equilibrium} allows us to derive an explicit form of the bound, the fluctuations of the velocity are in general not Gaussian and thus this result cannot be applied.
However, we can obtain a similar bound for a more general type of velocity statistics, which may be written as
\begin{align}
p_v(\bm{v} \vert \bm{x}) = q_v\big(\bm{v} - \bm{\nu}(\bm{x})\big) \label{shift-density},
\end{align}
with some arbitrary probability density $q_v(\bm{u})$.
This form corresponds to the case where only the first cumulant of the conditional velocity density depends on the position and all higher-order cumulants are independent of $\bm{x}$.
Just as in the case of local equilibrium dynamics, we find \eqref{potential-condition-local-equilibrium}.
Unfortunately, we cannot obtain an explicit solution to this equation for general $q_v(\bm{u})$.
However, we can easily find a sub-optimal solution if we do not impose the gradient condition and use \eqref{h-equation-4},
\begin{align}
&\grad_u \cdot \Big( \big(\bm{k}(\bm{x},\bm{u}) - \bm{J}_\nu(\bm{x}) \bm{u} \big) q_v(\bm{u}) \Big) = 0 .
\end{align}
This is obviously solved by
\begin{align}
\bm{k}(\bm{x},\bm{u}) = \bm{J}_\nu(\bm{x}) \bm{u} \label{h-first-cumulant} ,
\end{align}
which cannot be written as a velocity-gradient unless the Jacobian is symmetric.
Nevertheless, we obtain the bound
\begin{align}
2 \eta_J &\leq t \sigma^\text{od} + \frac{m^2 t}{\gamma T} \Av{\big\Vert \bm{J}_\nu \big( \bm{v} - \bm{\nu} \big) \big\Vert^2}  \\
& + \frac{4 m^2 t}{\gamma T} \Av{\big\Vert \bm{\alpha} \big\Vert^2} + 2 \Av{\big(\bm{\nu} \cdot \grad_v \ln p_v \big)^2} \n .
\end{align}
This can be brought into a form resembling \eqref{generalized-TUR-local} by expressing the covariance matrix of the velocity as
\begin{align}
\int d\bm{v} \ \big(v_i - \bm{\nu}_i(\bm{x}) \big) \big(v_j - \bm{\nu}_j(\bm{x}) \big) p_v(\bm{v} \vert \bm{x}) = \frac{T}{m} \Theta_{ij}  \label{theta},
\end{align}
where $\bm{\Theta}$ is a positive definite matrix independent of $\bm{x}$, which quantifies how much the local velocity fluctuations deviate from thermal fluctuations.
For the specific case \eqref{local-equilibrium}, $\bm{\Theta}$ is the identity matrix.
Using this, we obtain \eqref{main-bound-1-first-cumulant},
\begin{align}
2 \eta_J &\leq t \sigma^\text{od} + \frac{m t}{\gamma} \Av{\text{tr}\big(\bm{J}_\nu \bm{\Theta} \bm{J}_\nu^\text{T} \big)} \label{generalized-TUR-first-cumulant} \\
& + \frac{4 m^2 t}{\gamma T} \Av{\big\Vert \bm{\alpha} \big\Vert^2} + 2 \Av{\big(\bm{\nu} \cdot \grad_v \ln p_v \big)^2} \n .
\end{align}
Comparing this to \eqref{generalized-TUR-local}, the main difference is in the second term, where we account for the fact that the kinetic temperature defined by \eqref{theta} may deviate from the physical temperature $T$ and may be anisotropic.
Another difference is that this term includes the Jacobian instead of the symmetrized Jacobian.
As a consequence \eqref{generalized-TUR-first-cumulant} does not immediately reduce to \eqref{generalized-TUR-local} for the local equilibrium statistics \eqref{local-equilibrium}.
We have the relation
\begin{align}
\text{tr}\big(\bm{J}_\nu \bm{J}_\nu^\text{T}\big) = \text{tr}\big(\bm{J}_\nu^S \bm{J}_\nu^S \big) + \text{tr}\big(\bm{J}_\nu^A (\bm{J}_\nu^A)^\text{T} \big),
\end{align}
where $\bm{J}^A = (\bm{J}_\nu - \bm{J}_\nu^\text{T})/2$ is the antisymmetric part of the Jacobian.
Since both terms are positive, \eqref{generalized-TUR-first-cumulant} is less tight than \eqref{generalized-TUR-local}, since \eqref{h-first-cumulant} is not a gradient.
The above scheme can be extended to include higher-order cumulants.
For example, if only the first and second cumulant of the velocity density depend on the position, we may write its cumulant generating function as
\begin{align}
K_v(\bm{\lambda} \vert \bm{x}) &= \ln \bigg( \int d\bm{u} \ e^{\bm{\lambda} \cdot \bm{u}} q_v(\bm{u}\vert \bm{x}) \bigg) \label{cgf} \\
&\simeq \frac{T}{2 m} \bm{\lambda} \cdot \bm{\Theta}(\bm{x}) \bm{\lambda} + O(\lambda^3) \n ,
\end{align}
where the higher-order terms are independent of $\bm{x}$.
Here we again used the definition \eqref{theta} of the matrix $\bm{\Theta}(\bm{x})$, which may now depend on the position.
The velocity density can be expressed as the inverse Laplace transform of the exponential of the cumulant generating function
\begin{align}
q_v(\bm{u} \vert \bm{x}) = \mathbb{L}^{-1} \bigg( e^{K_v(\bm{\lambda} \vert \bm{x})} \bigg),
\end{align}
and thus its gradient with respect to $\bm{x}$ is
\begin{align}
\grad_x q_v(\bm{u} \vert \bm{x}) = \mathbb{L}^{-1} \bigg( e^{K_v(\bm{\lambda} \vert \bm{x})} \grad_x K_v(\bm{\lambda} \vert \bm{x})) \bigg) .
\end{align}
Using \eqref{cgf} and the properties of the Laplace transform with respect to differentiation, we find
\begin{align}
\bm{\nu}(\bm{x}) &\cdot \grad_x q_v(\bm{u}\vert \bm{x}) \\
&= \frac{T}{2 m} \grad_v \cdot \big[ (\bm{\nu}(\bm{x}) \cdot \grad_x) \bm{\Theta}(\bm{x}) \big] \grad_v q_v(\bm{u} \vert \bm{x}) \n .
\end{align}
This allows us to obtain the particular solution for $\bm{k}(\bm{x},\bm{u})$ from \eqref{h-equation-5},
\begin{align}
\bm{k}(\bm{x},\bm{u}) &= \bm{J}_\nu(\bm{x})\bm{u} \\
& \quad - \frac{T}{2 m} \big[(\bm{\nu}(\bm{x}) \cdot \grad_x) \bm{\Theta}(\bm{x}) \big] \grad_u \ln q_v(\bm{u}\vert \bm{x}) \n .
\end{align}
The resulting bound reads,
\begin{align}
2 \eta_J &\leq t \sigma^\text{od} + \frac{m t}{\gamma} \Av{\text{tr}\big(\bm{J}_\nu \bm{\Theta} \bm{J}_\nu^\text{T} \big)}  \label{generalized-TUR-second-cumulant} \\
& + \frac{4 m^2 t}{\gamma T} \Av{\big\Vert \bm{\alpha} \big\Vert^2}  + 2 \Av{\big(\bm{\nu} \cdot \grad_v \ln p_v \big)^2} \nn
& + \frac{m t}{\gamma} \Av{\text{tr}\big( \bm{J}_\nu^\text{T} (\bm{\nu} \cdot \grad_x) \bm{\Theta} \big)} \nn
& + \frac{T}{4 \gamma} \Av{ \grad_v \ln p_v \cdot \big[(\bm{\nu} \cdot \grad_x) \bm{\Theta} \big]^2 \grad_v \ln p_v}  \n .
\end{align}
Compared to \eqref{generalized-TUR-first-cumulant}, this includes two additional terms which account for the position-dependence of the kinetic temperature characterized by $\bm{\Theta}(\bm{x})$.

\subsection{Equilibrium fluctuation bound} \label{sec-bounds-equilibrium}
The bounds we discussed so far all involve the quantity $\eta_J$, that is, the ratio between the average current and its fluctuations, evaluated in the dynamics \eqref{langevin}.
As it turns out, we can derive a simpler bound relating the average current to its fluctuations, evaluated in the equilibrium dynamics \eqref{langevin-eq}.
The latter corresponds to the Kramers-Fokker-Planck equation
\begin{align}
\partial_t p(\bm{x},\bm{v},t) &= \mathcal{L}^\text{eq}(\bm{x},\bm{v}) p(\bm{x},\bm{v},t) \quad \text{with} \quad \label{kfp-eq}\\
\mathcal{L}^\text{eq}(\bm{x},\bm{v}) &= -\bm{v} \cdot \grad_x - \frac{1}{m} \grad_v \cdot \Big( \bm{\Phi}(\bm{x}) - \gamma \bm{v} - \frac{\gamma T}{m} \grad_v \Big) \n ,
\end{align}
with the force
\begin{align}
\bm{\Phi}(\bm{x}) = T \grad_x \ln p_x(\bm{x}) \label{equilibrium-force} .
\end{align}
It is straightforward to see that the steady state solution of this equation has the same steady state position probability density $p_x(\bm{x})$, but satisfies detailed balance with the equilibrium density \eqref{equilibrium-density} with $U(\bm{x}) = - T \ln p_x(\bm{x})$.
While \eqref{kfp-eq} has a vanishing steady state local mean velocity, we may still ask for an infinitesimal perturbation force $\epsilon \bm{f}(\bm{x},\bm{v})$ that, when applied to the dynamics \eqref{kfp-eq}, results in the leading order change to the steady state probability density
\begin{align}
\pi(\bm{x},\bm{v}) = -\bm{\nu}(\bm{x}) \cdot \grad_v p^\text{eq}(\bm{x},\bm{v}).
\end{align}
Intuitively, whereas above, we were looking for a perturbation that changes the local mean velocity from $\bm{\nu}(\bm{x})$ to $(1+\epsilon) \bm{\nu}(\bm{x})$, we are now looking for a perturbation that changes it from $0$ to $\epsilon \bm{\nu}(\bm{x})$.
However, in both cases the change in the current satisfies \eqref{current-change}.
The remaining argument proceeds exactly as before, however, we now replace $p(\bm{x},\bm{v})$ by $p^\text{eq}(\bm{x},\bm{v})$.
In particular, instead of \eqref{h-equation-2}, we now find
\begin{align}
\grad_v \cdot \Big( \big(\bm{h}(\bm{x},\bm{v}) - \bm{J}_\nu(\bm{x}) \bm{v} \big) p^\text{eq}_v(\bm{v}) \Big) &= 0 \label{h-equation-eq}.
\end{align}
In this case, since $p^\text{eq}_v(\bm{v})$ corresponds to a vanishing local mean velocity, the same is true for the local mean acceleration and the corresponding term is absent.
Imposing the potential condition on $\bm{h}(\bm{x},\bm{v})$ yields (see \eqref{potential-solution-local-equilibrium})
\begin{align}
\bm{h}(\bm{x},\bm{v}) = \frac{1}{2} \grad_v \big( \bm{v} \cdot \bm{J}_\nu^S(\bm{x}) \bm{v} \big) =  \bm{J}_\nu^S(\bm{x}) \bm{v} .
\end{align}
Plugging this into \eqref{generalized-TUR}, we then obtain \eqref{main-bound-2},
\begin{align}
2 \eta_J^\text{eq} \leq \bigg( t + \frac{2 m}{\gamma} \bigg) \sigma^\text{od} + \frac{m t}{\gamma} \Av{\text{tr}\Big(\big(\bm{J}_\nu^S\big)^2\Big)} \label{generalized-TUR-eq},
\end{align}
which is our second main result.
The advantage of this bound is that the right-hand side is now fully explicit and independent of the detailed velocity statistics of \eqref{langevin}.
Interestingly, the right-hand side is exactly the same as the leading-order term in the small mass limit, \eqref{main-bound-1-over}.
The tradeoff is that the variance $\text{Var}(J(t))$ is replaced by the variance $\text{Var}^\text{eq}(J(t))$ in the equilibrium dynamics \eqref{kfp-eq}, which cannot be determined directly from \eqref{langevin}.
However, in most cases, we expect that the fluctuations are enhanced by driving the system out of equilibrium.
In such situations, we have $\text{Var}^\text{eq}(J(t)) \leq \text{Var}(J(t))$, and thus \eqref{generalized-TUR-eq} also provides a bound on the precision of the current in the original dynamics.

\section{Magnetic field} \label{sec-magnetic}
The same formalism that we used in the previous section may also be employed to derive bounds on the precision of a current in the presence of a Lorentz force due to a magnetic field.
In this case, the Kramers-Fokker-Plank equation reads
\begin{align}
\partial_t p(\bm{x},\bm{v},t) &= \mathcal{L}(\bm{x},\bm{v}) p(\bm{x},\bm{v},t) \qquad \text{with} \label{kfp-magnetic}\\
\mathcal{L}(\bm{x},\bm{v}) &= -\bm{v} \cdot \grad_x - \frac{1}{m} \grad_v \cdot \Big( \bm{F}(\bm{x}) + q \bm{v} \times \bm{B}(\bm{x}) \nn
&\hspace{2cm} - \gamma \bm{v} - \frac{\gamma T}{m} \grad_v \Big) \n ,
\end{align}
where $\bm{B}(\bm{x})$ is an external magnetic field and $q$ is the charge of the particle.
For this case, it has been shown previously that the TUR \eqref{TUR} can be violated \cite{Chu19}; indeed, the left-hand side can become arbitrarily large since a strong magnetic field can suppress the fluctuations of the current, while the entropy production remains finite.
As before, we want to find an infinitesimal force $\epsilon \bm{f}(\bm{x},\bm{v})$ that leads to a rescaling of the local mean velocity via the first-order correction of the probability density
\begin{align}
\pi(\bm{x},\bm{v}) = -\bm{\nu}(\bm{x}) \cdot \grad_v p(\bm{x},\bm{v}).
\end{align}
For \eqref{kfp-magnetic}, the force is given by (see Appendix \ref{app-magnetic-force} for the derivation)
\begin{align}
\bm{f}(\bm{x},\bm{v}) = \gamma \bm{\nu}(\bm{x}) - q \bm{\nu}(\bm{x}) \times \bm{B}(\bm{x}) + m \bm{h}(\bm{x},\bm{v}) \label{force-magnetic},
\end{align}
where $\bm{h}(\bm{x},\bm{v})$ is determined by the equation
\begin{align}
\grad_v \cdot \Big( \bm{h}(\bm{x},\bm{v}) p_v(\bm{v}\vert \bm{x}) \Big) &= \grad_v \cdot \Big( p_v(\bm{v}\vert \bm{x}) (\bm{v} \cdot \grad_x ) \bm{\nu}(\bm{x})\Big) \nn
& \quad - \bm{\nu}(\bm{x}) \cdot \grad_x p_v(\bm{v}\vert \bm{x}) \label{h-equation-magnetic},
\end{align}
which is the same as \eqref{h-equation-2}.
Surprisingly, the magnetic field does not enter the equation determining $\bm{h}(\bm{x},\bm{v})$ explicitly, even though it appears implicitly via the form of the conditional velocity probability density $p_v(\bm{v}\vert \bm{x})$, which may depend on the magnetic field.
Thus, the local mean value of $\bm{h}(\bm{x},\bm{v})$ is still given by twice the local mean acceleration $\bar{\bm{h}}(\bm{x}) = 2 \bm{\alpha}(\bm{x})$.
The next step is to evaluate the right-hand side of \eqref{fri-bound} for the perturbation force \eqref{force-magnetic}.
We note that the first and second term of \eqref{force-magnetic} are orthogonal, $\bm{\nu}(\bm{x}) \cdot (\bm{\nu}(\bm{x}) \times \bm{B}(\bm{x})) = 0$.
Thus, \eqref{fri-bound} yields
\begin{subequations}
\begin{align}
\Psi &= \frac{4 m^2}{\gamma T}  \Av{ \big\Vert \bm{\alpha} \big\Vert^2} + \frac{m^2}{\gamma T} \Av{ \big\Vert \tilde{\bm{h}} \big\Vert^2}  \\
& \qquad + \frac{q^2}{T \gamma} \Av{\big\Vert \bm{\nu} \times \bm{B} \big\Vert^2} - \frac{2 m q}{\gamma T} \Av{\bm{h} \cdot \big( \bm{\nu} \times \bm{B} \big)}  \nn 
\Omega_\nu &= 2 \Av{ \big(\bm{\nu} \cdot \grad_v \ln p_v \big)^2} .
\end{align}
\end{subequations}
In the fourth term, we can replace $\bm{h}(\bm{x},\bm{v})$ by its local mean value, since the second factor does not depend on $\bm{v}$.
Defining the local mean Lorentz force as $\bar{\bm{F}}^\text{L}(\bm{x}) =q\bm{\nu}(\bm{x}) \times \bm{B}(\bm{x})$, we obtain
\begin{subequations}
\begin{align}
\Psi &= \frac{1}{\gamma T}  \Av{ \big\Vert 2 m \bm{\alpha} - \bar{\bm{F}}^\text{L} \big\Vert^2} + \frac{m^2}{\gamma T} \Av{ \big\Vert \grad_v \psi \big\Vert^2}  \label{generalized-TUR-magnetic} \\
\Omega_\nu &= 2 \Av{ \big(\bm{\nu} \cdot \grad_v \ln p_v \big)^2} .
\end{align}
\end{subequations}
As before, we minimized the right-hand side with respect to $\bm{h}(\bm{x},\bm{v})$, which results in the condition that $\bm{h}(\bm{x},\bm{v})$ is the velocity-gradient of a scalar function.
Compared to \eqref{main-bound-1}, we see that the value of the local mean acceleration is shifted by the local mean Lorentz force.
In the overdamped limit, the latter term dominates and we obtain
\begin{align}
&2 \eta_J \leq t \sigma^\text{od} + \frac{t }{\gamma T } \Av{\big\Vert \bar{\bm{F}}^\text{L} \big\Vert^2} + O(m) \label{generalized-TUR-over-magnetic} .
\end{align}
Thus, we acquire an additional terms which is proportional to the square of the local mean Lorentz force.
If we have a constant magnetic field and the local mean velocity is orthogonal to it, then $\Vert \bar{\bm{F}}^L(\bm{x}) \Vert^2 = q^2 \Vert \bm{\nu}(\bm{x}) \Vert^2 \Vert \bm{B} \Vert^2$, and we can write \eqref{generalized-TUR-over-magnetic}
as
\begin{align}
2 \eta_J \leq \bigg( 1 + \frac{q^2 \Vert \bm{B} \Vert^2}{\gamma^2}\bigg) t \sigma^\text{od} .
\end{align}
This is precisely the overdamped TUR in the presence of a magnetic field derived in Ref.~\cite{Par21}, where the factor in front of the entropy production was interpreted as a reduced effective temperature.
As before, we can obtain more explicit forms of \eqref{generalized-TUR-magnetic} using suitable assumptions on the velocity statistics.
If only the first cumulant depends on the position, then we find, in analogy to \eqref{generalized-TUR-first-cumulant},
\begin{align}
\Psi &= \frac{m}{\gamma} \Av{\text{tr}\big(\bm{J}_\nu \bm{\Theta} \bm{J}_\nu^\text{T} \big)} + \frac{1}{\gamma T}  \Av{ \big\Vert 2 m \bm{\alpha} - \bar{\bm{F}}^\text{L} \big\Vert^2}   \label{generalized-TUR-first-magnetic}  .
\end{align}
If the relative velocity statistics are thermal as in \eqref{local-equilibrium}, then this further simplifies to
\begin{align}
\Psi &= \frac{m}{\gamma} \Av{\text{tr}\Big( \big(\bm{J}_\nu^S \big)^2 \Big)} + \frac{1}{\gamma T}  \Av{ \big\Vert 2 m \bm{\alpha} - \bar{\bm{F}}^\text{L} \big\Vert^2}  \label{generalized-TUR-local-magnetic} .
\end{align}
Finally, since the magnetic field does not change the Boltzmann-Gibbs density \eqref{equilibrium-density}, the derivation of the equilibrium fluctuation bound \eqref{generalized-TUR-eq} is completely analog to the case without a magnetic field and we obtain
\begin{align}
2 \eta_J^\text{eq} &\leq \Big(t + 2 \frac{m}{\gamma} \Big) \sigma^\text{od} + \frac{m t}{\gamma} \Av{\text{tr}\Big( \big(\bm{J}_\nu^S \big)^2 \Big)} \label{generalized-TUR-eq-magnetic} \\
&\qquad + \frac{t}{\gamma T} \Av{ \big\Vert \bar{\bm{F}}^\text{L} \big\Vert^2}   \n  ,
\end{align}
which corresponds to setting the local mean acceleration to zero in \eqref{generalized-TUR-local-magnetic}.

\section{Anisotropic and coordinate-dependent temperature and friction} \label{sec-aniso}
All relations derived in the previous sections generalize in a straightforward manner to the case where the temperature and friction are anisotropic and/or depend on the coordinate $\bm{x}$.
Under these conditions, the constants $\gamma$ and $T$ in \eqref{kfp} are replaced by positive definite coordinate-dependent matrices $\bm{\gamma}(\bm{x})$ and $\bm{T}(\bm{x})$.
For simplicity, we focus on the case where both matrices are diagonal and thus commute.
We can also allow for the mass to be a positive diagonal matrix $\bm{m}$.
This describes, for example, the case of interacting particles, whose masses and sizes are different and which may be in contact with different heat baths.
The entropy production rate \eqref{entropy-splitting} can then be written as
\begin{align}
\sigma &= \Av{ \bm{\nu} \cdot \bm{\gamma} \bm{T}^{-1} \bm{\nu}} \\
& \quad + \Av{\big(\bm{v} - \bm{\nu} \big) \cdot \bm{\gamma} \bm{T}^{-1} \big(\bm{v} - \bm{\nu} \big)} - \Av{ \text{tr} \big( \bm{\gamma} \bm{m}^{-1} \big)} \n ,
\end{align}
where, as before, we identify the term in the first line as the overdamped entropy production rate $\sigma^\text{od}$.
As in Sec.~\ref{sec-bounds-perturbation}, we want to find a perturbation force that results in the leading order change in the probability density \eqref{probability-mod}.
This force is given by
\begin{align}
\bm{f}(\bm{x},\bm{v}) = \bm{\gamma} \bm{\nu}(\bm{x}) + 2 \bm{m} \bm{\alpha}(\bm{x}) + \bm{m} \grad_v \psi(\bm{x},\bm{v}),
\end{align}
where $\psi(\bm{x},\bm{v})$ still has to obey \eqref{psi-equation-main}.
The bound corresponding to \eqref{fri-bound} is given by
\begin{align}
2 \eta_J \leq t \Av{\bm{f} \cdot \big(\bm{\gamma} \bm{T} \big)^{-1} \bm{f}} + 2 \Av{\big(\bm{\nu} \cdot \grad_v \ln p_v \big)^2} .
\end{align}
Since the temperature, friction and mass are now matrix-valued, we cannot use \eqref{orthogonality-h}.
Instead of \eqref{main-bound-1}, we now obtain
\begin{align}
2 \eta_J &\leq t \Av{\big( \bm{\gamma} \bm{\nu} + 2 \bm{m} \bm{\alpha} \big) \cdot \big(\bm{\gamma} \bm{T} \big)^{-1} \big( \bm{\gamma} \bm{\nu} + 2 \bm{m} \bm{\alpha} \big)} \\
& \qquad + t \Av{ \grad_v \psi \cdot \bm{m}^2 \big(\bm{\gamma} \bm{T} \big)^{-1} \grad_v \psi} \nn
& \qquad + 2 \Av{\big(\bm{\nu} \cdot \grad_v \ln p_v \big)^2} \n .
\end{align}
While the first term is positive, the contributions from the local mean velocity and the local mean acceleration no longer separate into two explicitly positive terms.
This means that we cannot rule out that the first term may be smaller than the overdamped entropy production rate, and thus the bound may in principle be tighter than the overdamped TUR.
We leave the question of whether this can be realized by a concrete system to future research.

\section{Multidimensional TUR and correlations} \label{sec-multi-tur}
In Ref.~\cite{Dec18c}, the overdamped steady-state TUR was generalized to a joint bound on several different currents.
We can extend this generalization to the results derived in Sec.~\ref{sec-bounds} by noting that \eqref{fri-bound} is essentially the Cram{\'e}r-Rao inequality for the path probability density,
\begin{align}
\frac{\big(\partial_\epsilon \av{J}^\epsilon \vert_{\epsilon = 0} \big)^2}{\text{Var}(J)} \leq I,
\end{align}
where the right-hand side is the Fisher information of the path probability density with respect to the parameter $\epsilon$, which is given by the right-hand side of \eqref{fri-bound}.
If we consider several currents $\bm{J}(t) = (J_1(t),\ldots,J_K(t))$, then this inequality generalizes to
\begin{align}
\big(\partial_\epsilon \av{\bm{J}(t)}^\epsilon \vert_{\epsilon = 0} \big) \cdot \big(\bm{\Xi}_J \big)^{-1} \big(\partial_\epsilon \av{\bm{J}(t)}^\epsilon \vert_{\epsilon = 0} \big) \leq I \label{cramer-rao},
\end{align}
where we introduced the covariance matrix
\begin{align}
\big(\bm{\Xi}_J \big)_{ij} &= \text{Cov}(J_i(t),J_j(t)) \\
&= \av{J_i(t) J_j(t)} - \av{J_i(t)} \av{J_j(t)} . \n
\end{align}
Using this, \eqref{main-bound-1} readily generalizes to a joint bound on several currents,
\begin{align}
\av{\bm{J}(t)} \cdot \big(\bm{\Xi}_J\big)^{-1} \av{\bm{J}(t)} \leq \frac{1}{2} \Sigma ,
\end{align}
where $\Sigma$ has the same form as in \eqref{main-bound-1}.
Similarly, we can generalize \eqref{main-bound-2} using the covariance matrix of the currents in the equilibrium dynamics \eqref{langevin-eq},
\begin{align}
\av{\bm{J}(t)} \cdot \big(\bm{\Xi}^\text{eq}_J\big)^{-1} \av{\bm{J}(t)} \leq \frac{1}{2} \Sigma^\text{eq} .
\end{align}
Further, since the perturbation \eqref{force-h} does not change the position density $p_x(\bm{x})$, the average of observables that only depend on the position is invariant under the perturbation.
In particular, for time-integrated observables of the type
\begin{align}
Z(t) = \int_0^t ds \ z(x(s)) \label{position-observable},
\end{align}
we have
\begin{align}
\av{Z(t)}^\epsilon \simeq \av{Z(t)}^{\epsilon = 0} = t \av{z} .
\end{align}
For such observables, we then have
\begin{align}
\partial_\epsilon \av{Z(t)}^\epsilon \vert_{\epsilon = 0} = 0.
\end{align}
Following Ref.~\cite{Dec21}, we can consider the joint bound of a current $J(t)$ and a time-integrated, state-dependent observable $Z(t)$,
\begin{align}
\eta_{J,Z} \equiv \frac{\av{J(t)}^2}{\text{Var}(J(t)) - \frac{\text{Cov}(J(t),Z(t))^2}{\text{Var}(Z(t))}} \leq \frac{1}{2} \Sigma \label{TUR-corr} .
\end{align}
Since the denominator is always smaller than the variance of $J(t)$, this inequality is tighter than \eqref{main-bound-1} for arbitrary $Z(t)$.
This bound has been termed correlation TUR (CTUR) in Ref.~\cite{Dec21}.
As argued there for the overdamped case, we can generally expect a relatively tight bound for the choice
\begin{align}
Z(t) = \bar{J}(t) = \int_0^t ds \ \bm{w}(\bm{x}(s)) \cdot \bm{\nu}(\bm{x}(s) ) \label{local-mean-current},
\end{align}
which is obtained by replacing the velocity by the local mean velocity in \eqref{current} and can be regarded as the local mean value of the current $J(t)$.

\section{Finite perturbations} \label{sec-finite-perturbation}
In the derivation of \eqref{main-bound-1} in Sec.~\ref{sec-bounds}, we considered a specific perturbation that leads to an infinitesimal rescaling of the local mean velocity to $(1+\epsilon) \bm{\nu}(\bm{x})$.
Here, we show that this perturbation actually arises as the leading order expression of a more general perturbation.
We again use the transformation \eqref{conditional-density-relative} to characterize the velocity statistics relative to the local mean value,
\begin{align}
q_v(\bm{u} \vert \bm{x}) = p_v(\bm{u} + \bm{\nu}(\bm{x}) \vert \bm{x}) .
\end{align}
\begin{widetext}
\noindent In terms of the shifted velocity $\bm{u} = \bm{v} - \bm{\nu}(\bm{x})$, the Kramers-Fokker-Planck equation can be written as
\begin{align}
0 = \Bigg( - \big(\bm{u} + \bm{\nu}(\bm{x}) \big) &\cdot \grad_x - \frac{1}{m} \grad_u \bigg( \bm{F}(\bm{x}) - \gamma \big(\bm{u} + \bm{\nu}(\bm{x}) \big) - \frac{\gamma T}{m} \grad_u \bigg)  \label{kfp-conditional} \\
& + \Big[\Big(\big(\bm{u} + \bm{\nu}(\bm{x}) \big) \cdot \grad_x \Big) \bm{\nu}(\bm{x})  \Big] \cdot \grad_u \Bigg) \big( q_v(\bm{u}\vert \bm{x}) p_x(\bm{x}) \big) \n .
\end{align}
We now want to add a force $\bm{f}_\epsilon(\bm{x},\bm{v})$ that leads to a rescaling of the local mean velocity, $(1+\epsilon) \bm{\nu}(\bm{x})$, while leaving the functional form of $q_v(\bm{u} \vert \bm{x})$ and $p_x(\bm{x})$ invariant.
In terms of the cumulants of $p_v(\bm{v}\vert \bm{x})$, this means that we want to change the overall size of the first cumulant while keeping the remaining cumulants the same.
Note that, in contrast to Sec.~\ref{sec-bounds}, we do not assume $\epsilon$ to be small.
Using this condition in \eqref{kfp-conditional}, we get
\begin{align}
0 = \Bigg( - \big(\bm{u} + (1+\epsilon)\bm{\nu}(\bm{x}) \big) &\cdot \grad_x - \frac{1}{m} \grad_u \bigg( \bm{F}(\bm{x}) + \bm{f}^\epsilon(\bm{x},\bm{v}) - \gamma \big(\bm{u} + (1+\epsilon)\bm{\nu}(\bm{x}) \big) - \frac{\gamma T}{m} \grad_u \bigg) \\
& + (1+\epsilon)\Big[\Big(\big(\bm{u} + (1+\epsilon)\bm{\nu}(\bm{x}) \big) \cdot \grad_x \Big) \bm{\nu}(\bm{x})  \Big] \cdot \grad_u \Bigg) \big( q_v(\bm{u}\vert \bm{x}) p_x(\bm{x}) \big) \n .
\end{align}
Since both this and \eqref{kfp-conditional} have to hold, we can subtract them and obtain
\begin{align}
0 = \Bigg( - \epsilon \bm{\nu}(\bm{x}) &\cdot \grad_x - \frac{1}{m} \grad_u \bigg( \bm{f}^\epsilon(\bm{x},\bm{v}) - \epsilon \gamma \bm{\nu}(\bm{x}) \bigg) \\
& + \epsilon \Big[\big(\bm{u} \cdot \grad_x \big) \bm{\nu}(\bm{x})  \Big] \cdot \grad_u + \big(2 \epsilon + \epsilon^2 \big) \Big[\big(\bm{\nu}(\bm{x}) \cdot \grad_x \big) \bm{\nu}(\bm{x})  \Big] \cdot \grad_u \Bigg) \big( q_v(\bm{u}\vert \bm{x}) p_x(\bm{x}) \big) \n .
\end{align}
Using \eqref{stationary-condition} and rearranging, we obtain an equation for the force $\bm{f}_\epsilon(\bm{x},\bm{v})$,
\begin{align}
\grad_u \cdot \bigg( \Big( \bm{f}^\epsilon(\bm{x},\bm{v}) - \epsilon \gamma \bm{\nu}(\bm{x}) - m \big(2 \epsilon + \epsilon^2) \bm{J}_\nu(\bm{x}) \bm{\nu}(\bm{x}) - m \epsilon \bm{J}_\nu(\bm{x}) \bm{u} \Big) q_v(\bm{u} \vert \bm{x}) \bigg) = - m \epsilon \bm{\nu}(\bm{x}) \cdot \grad_x q_v(\bm{u}\vert \bm{x}) .
\end{align}
Writing the force as
\begin{align}
\bm{f}^\epsilon(\bm{x},\bm{v}) = \epsilon \gamma \bm{\nu}(\bm{x}) + m \big(2 \epsilon + \epsilon^2) \bm{\alpha}(\bm{x}) + m \epsilon \bm{k}(\bm{x},\bm{v}) \label{force-finite},
\end{align}
we obtain precisely \eqref{h-equation-4},
\begin{align}
&\grad_u \cdot \Big( \bm{k}(\bm{x},\bm{u}) q_v(\bm{u}\vert \bm{x}) \Big) = \grad_u \cdot \Big( \bm{J}_\nu(\bm{x}) \bm{u} q_v(\bm{u}\vert \bm{x}) \Big) - \bm{\nu}(\bm{x}) \cdot \grad_x q_v(\bm{u}\vert \bm{x}) .
\end{align}
\end{widetext}
The only difference between the force constructed in Sec.~\ref{sec-bounds-perturbation} and \eqref{force-finite} is an additional term proportional to $\epsilon^2$ time the local mean acceleration.
In particular, we recover the previous result as the leading order perturbation force in the limit $\epsilon \rightarrow 0$.
Further, the perturbation of the probability density \eqref{probability-mod}, which was previously chosen ad-hoc as a particular choice that produces the correct scaling of the current, follows naturally from
\begin{align}
p_v^\epsilon(\bm{v} \vert \bm{x}) &= q_v(\bm{v} - (1+\epsilon) \bm{\nu}(\bm{x}) \vert \bm{x})  \\
&\simeq q_v(\bm{v} - \bm{\nu}(\bm{x}) \vert \bm{x}) - \epsilon \bm{\nu}(\bm{x}) \cdot \grad_v q_v(\bm{v} - \bm{\nu}(\bm{x}) \vert \bm{x}) \nn
&= p(\bm{v} \vert \bm{x}) - \epsilon \bm{\nu}(\bm{x}) \cdot \grad_v p_v(\bm{v} \vert \bm{x}) \n ,
\end{align}
in the limit of small $\epsilon$.
Thus, \eqref{probability-mod} indeed corresponds to the first-order correction to the probability density for a infinitesimal rescaling of the local mean velocity.
We stress that this detailed control over the statistics of the system---changing the local mean velocity while leaving the shape of the velocity density invariant---requires the force to depend on both position and velocity.
As a consequence, $\bm{f}^\epsilon(\bm{x},\bm{v})$ is not a force that is realizable in practice; the fact that physical forces typically only depend on the position of the particle means that underdamped dynamics cannot be fully controlled in general \cite{Mur14,Mur14b,Bau16}.
However, for the purposes of deriving bounds on the precision of the current, the existence of such a force is sufficient.

\section{Examples} \label{sec-examples}

\subsection{Biased diffusion} \label{sec-examples-diffusion}
\begin{figure}
\includegraphics[width=.47\textwidth]{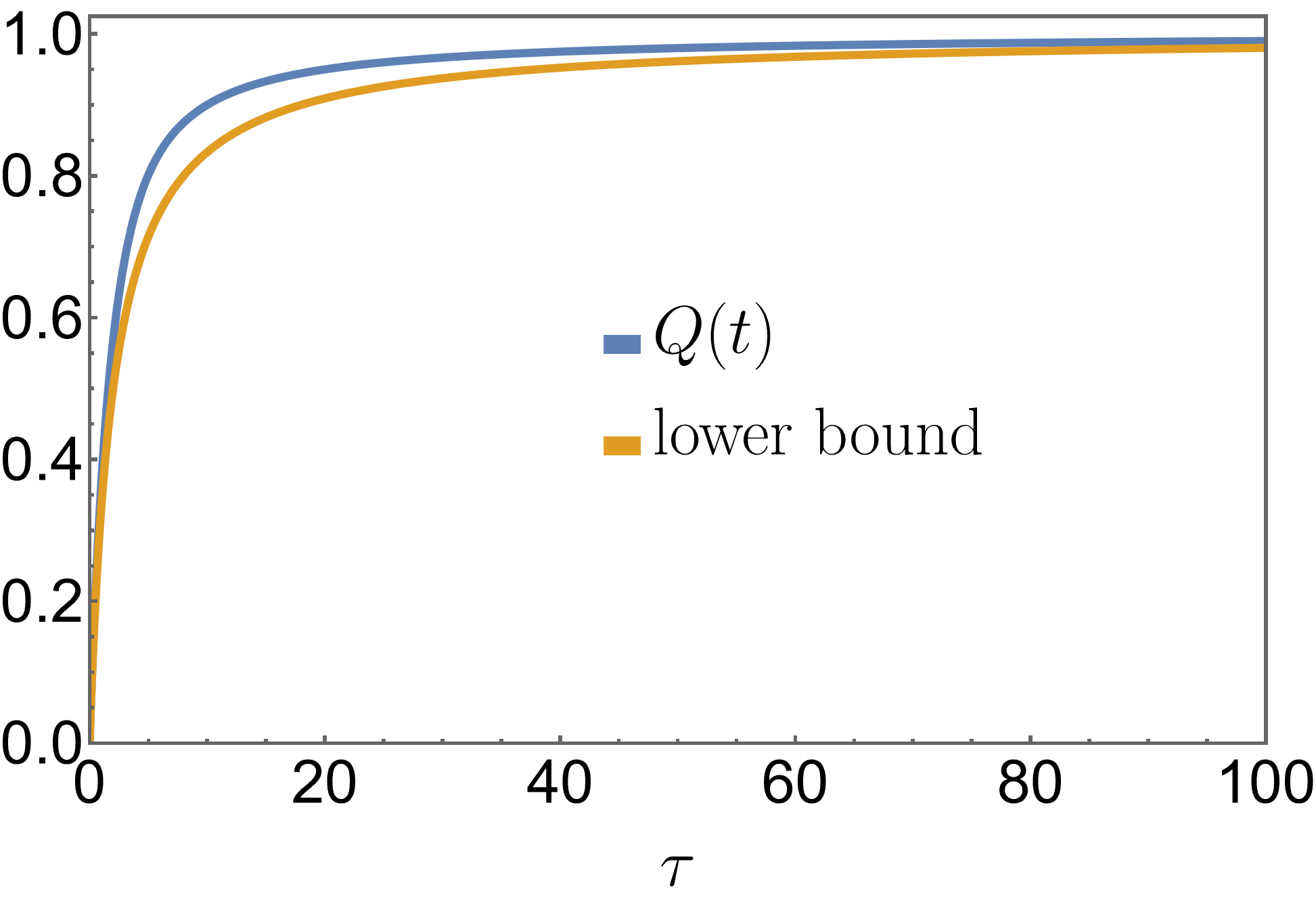}
\caption{The uncertainty product \eqref{uncertainty-diffusion} (blue) as a function of the dimensionless time $\tau = t \gamma/m$. The orange line is the lower bound \eqref{bound-diffusion}, which coincides with the actual value both at short times and at long times, where both approach the overdamped limiting value of $1$. }
\label{fig-diffusion}
\end{figure}

The most simple example of a non-equilibrium dynamics of the form \eqref{langevin} is when the force $\bm{F}$ is constant, and we consider an arbitrary periodic domain of volume $V$.
In this case, the steady state probability density is given by
\begin{align}
p(\bm{x},\bm{v}) = \frac{1}{V} \bigg(\sqrt{\frac{m}{2 \pi T}} \bigg)^3 \exp\bigg[ - \frac{m\big\Vert \bm{v} - \frac{1}{\gamma} \bm{F} \big\Vert^2}{2 T} \bigg].
\end{align}
Formally, this belongs to the class of local equilibrium dynamics discussed in \ref{sec-bounds-local}, with the additional simplification that the local mean velocity is constant and, thus, the local mean acceleration vanishes.
We then obtain from \eqref{main-bound-1-local},
\begin{align}
&2 \eta_J \leq \Big(t + 2 \frac{m}{\gamma} \Big) \sigma \label{bound-diffusion} .
\end{align}
In \cite{Fis20}, the uncertainty product
\begin{align}
Q_J(t) \equiv \frac{\sigma}{2 \eta_J} \label{uncertainty}
\end{align}
for this system for the displacement $z(t) = \int_0^t ds \ v(s)$ was computed explicitly as
\begin{align}
Q_z(t) = \frac{1}{\tau}\Big(\tau - 1 + e^{-\tau} \Big) \quad \text{with} \quad \tau = \frac{t \gamma}{m} \label{uncertainty-diffusion} ,
\end{align}
whereas the bound \eqref{bound-diffusion} yields
\begin{align}
Q_z(t) \geq \frac{1}{1 + \frac{2}{\tau}} .
\end{align}
The lower bound is sharp both at short and long times, with a relative deviation of less than $14 \%$ at intermediate times, see Fig.~\ref{fig-diffusion}.

\subsection{Biased diffusion in a uniform magnetic field} \label{sec-examples-diffusion-magnetic}
\begin{figure}
\includegraphics[width=.47\textwidth]{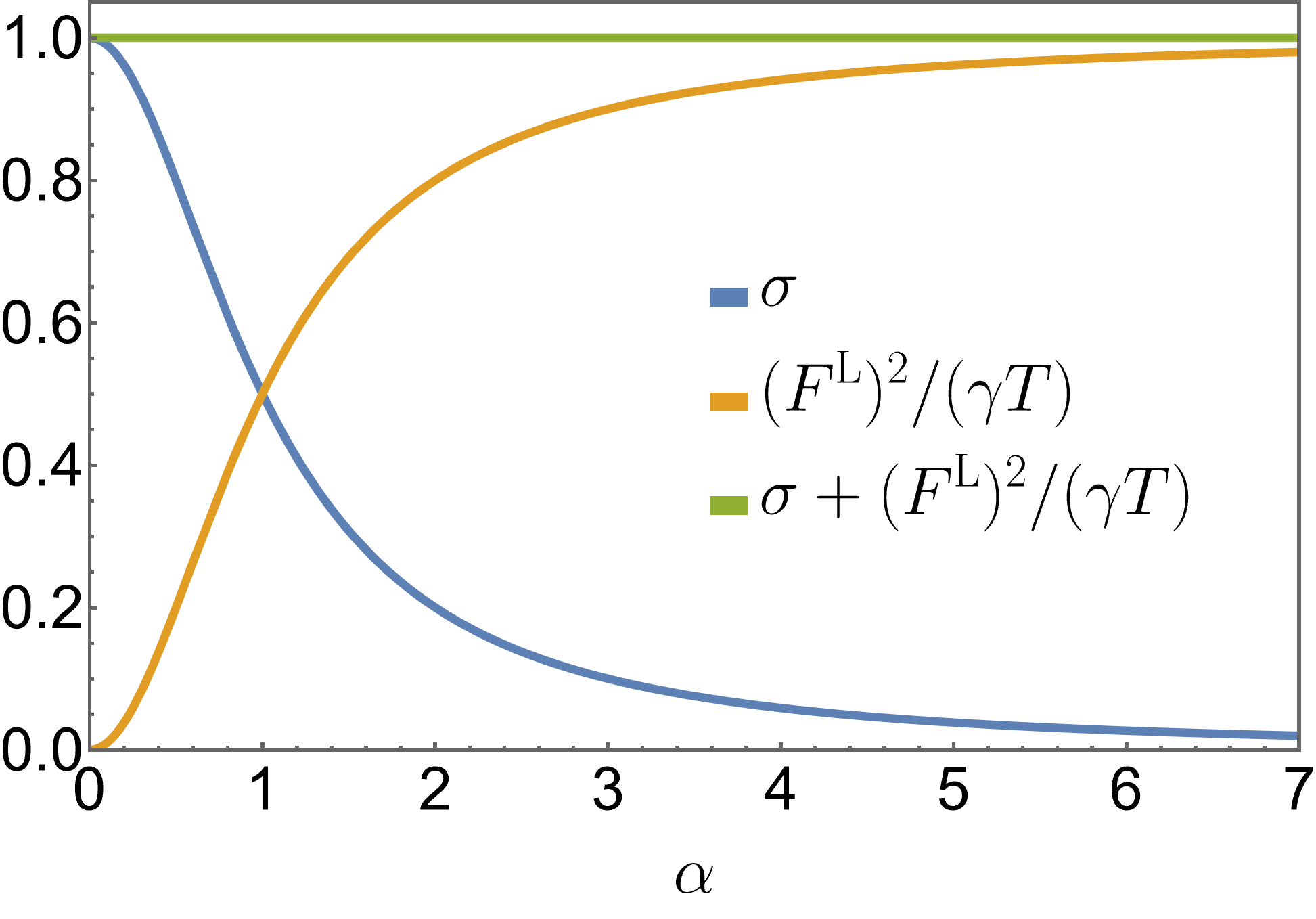}
\caption{The behavior of the first (blue) and second (orange) term in \eqref{bound-diffusion-magnetic} and their sum (green) as a function of $\alpha = q B/\gamma$ for $F^2/(\gamma T) = 1$. \label{fig-diffusion-magnetic}}
\end{figure}

\begin{figure}
\includegraphics[width=.47\textwidth]{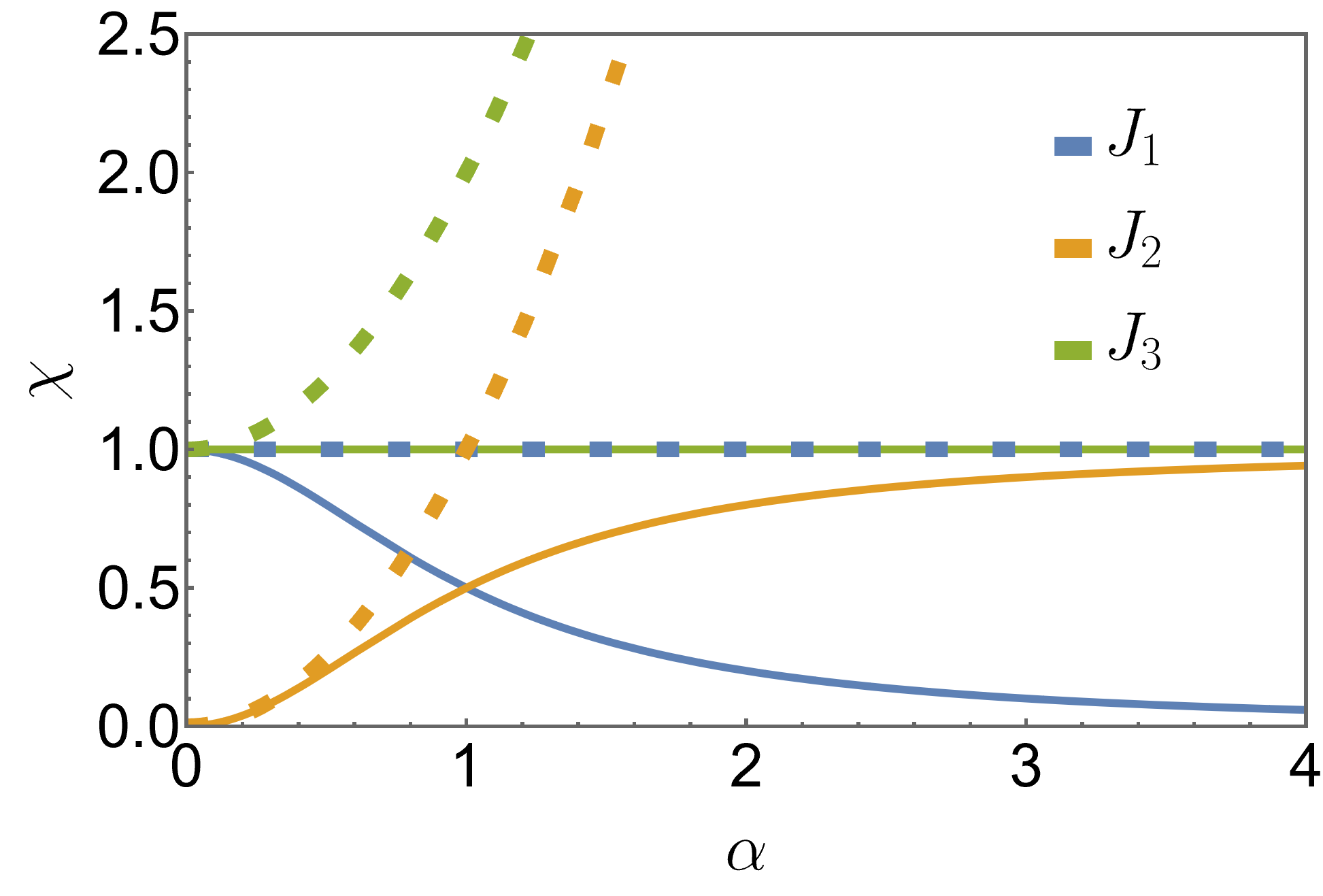}
\caption{The ratios $\chi_J$ (solid lines) and $\tilde{\chi}_J$ (dashed lines) defined in \eqref{uncertainty-magnetic}, for the three observables defined in \eqref{observables-magnetic} as a function of the strength of the magnetic field $\alpha = q B/\gamma$.}
\label{fig-diffusion-magnetic-2}
\end{figure}

A slight variation on the previous example is obtained by additionally including a uniform magnetic field, which, without loss of generality, we assume to be in the $x_3$-direction, $\bm{B} = (0,0,B)$.
In this case, the steady state probability density is obtained as
\begin{align}
p(\bm{x},\bm{v}) = \frac{1}{V} \bigg(\sqrt{\frac{m}{2 \pi T}}\bigg)^3 \exp\bigg[ - \frac{m\big\Vert \bm{v} - \bm{\mu} \bm{F} \big\Vert^2}{2 T} \bigg],
\end{align}
where we introduced the mobility matrix
\begin{align}
\bm{\mu} = \frac{1}{\gamma^2 + (q B)^2} \left(\begin{array}{ccc}
\gamma & q B & 0 \\[1 ex]
-q B & \gamma & 0 \\[1 ex]
0 & 0 & \gamma + \frac{(q B)^2}{\gamma}
\end{array} \right) .
\end{align}
As before, this is a local equilibrium dynamics with $\bm{\nu} = \bm{\mu} \bm{F}$, which is independent of $\bm{x}$.
From \eqref{generalized-TUR-local-magnetic}, we obtain
\begin{align}
2 \eta_J \leq \Big(t + 2 \frac{m}{\gamma} \Big) \sigma + \frac{t q^2}{T \gamma} \big\Vert (\bm{\mu} \bm{F}) \times \bm{B} \big\Vert^2 \label{bound-diffusion-magnetic}
\end{align}
Explicitly, the entropy production rate and the second term are given by
\begin{subequations}
\begin{align}
&\sigma = \frac{F^2}{\gamma  T (1 + \alpha^2)} \\
&\frac{q^2}{\gamma T} \big\Vert (\bm{\mu} \bm{F}) \times \bm{B} \big\Vert^2 = \frac{(F^\text{L})^2}{\gamma T} = \frac{F^2}{\gamma T}  \frac{\alpha^2 + \alpha^4}{(1 + \alpha^2)^2} ,
\end{align}
\end{subequations}
where $F^\text{L}$ is the magnitude of the Lorentz force.
For simplicity and without loss of generality, we here assumed the force to be in the $x_1$-direction, $\bm{F} = (F,0,0)$ and introduced the dimensionless parameter $\alpha = q B/\gamma$, which measures the strength of the magnetic field compared to the damping.
Both terms are proportional to $F^2$, their behavior as a function of $\alpha$ is shown in Fig.~\ref{fig-diffusion-magnetic}.
For $\alpha = 0$, the second term tends to zero and we recover \eqref{bound-diffusion}.
For $\alpha \gg 1$, on the other hand, the entropy production rate tends to zero and the second term tends to a constant.
Interestingly, the sum of both terms is same as the entropy production rate for $\alpha = 0$ (see Fig.~\ref{fig-diffusion-magnetic}).
This allows us to write \eqref{bound-diffusion-magnetic} as
\begin{align}
2 \eta_J \leq \Big(t (1+\alpha^2) + 2 \frac{m}{\gamma} \Big) \sigma .
\end{align}
In order to verify the bound \eqref{bound-diffusion-magnetic}, we consider the three observables
\begin{align}
J_1(t) &= \int_0^t ds \ v_1(s), \quad J_2(t) = \int_0^t ds \ v_2(s), \label{observables-magnetic} \\
J_3(t) &= \int_0^t ds \ \hat{\bm{\nu}} \cdot \bm{v}(s) \n  .
\end{align}
The first two observables are just the displacement in $x_1$ and $x_2$ direction, respectively, whereas the third one is the displacement in the direction $\hat{\bm{\nu}} = \bm{\nu}/\Vert \bm{\nu} \Vert$ of the local mean velocity.
Their steady-state averages are given by
\begin{align}
\av{J_1(t)} &= \frac{F t}{\gamma (1+\alpha^2)}, \quad \av{J_2(t)} = \frac{F \alpha t}{\gamma (1+\alpha^2)}, \\
\av{J_3(t)} &= \frac{F t}{\gamma \sqrt{1+\alpha^2}} \n .
\end{align}
For simplicity, we focus on the long-time limit, where the variance of all three observables is
\begin{align}
\text{Var}(J_{1,2,3}(t)) \simeq \frac{2 T}{\gamma (1+\alpha^2)} t,
\end{align}
since the diffusion in different spatial directions decouples in the long-time limit.
We see that, as noted in \cite{Chu19}, a non-zero magnetic field reduces the diffusion coefficient.
For each of these observables, we show the long-time limiting value of the ratios
\begin{align}
\chi_J &\equiv \lim_{t \rightarrow \infty} \frac{2 \av{J(t)}^2}{\text{Var}(J(t)) (1+\alpha^2) \sigma t} \leq 1, \label{uncertainty-magnetic} \\
\tilde{\chi}_J &\equiv \lim_{t \rightarrow \infty} \frac{2 \av{J(t)}^2}{\text{Var}(J(t)) \sigma t} \n ,
\end{align}
as a function of $\alpha$ in Fig.~\ref{fig-diffusion-magnetic-2}.
Whereas $\chi_J$ corresponds to the long-time behavior of \eqref{bound-diffusion-magnetic}, $\tilde{\chi}_J$ corresponds to the naive TUR \eqref{TUR} without the additional terms stemming from the magnetic field.
In the absence of a magnetic field, the displacement in $x_1$ direction (parallel to the applied force) saturates the upper bound $\chi_{J_1} = 1$; the system is then equivalent to the one discussed in Sec.~\ref{sec-examples-diffusion}.
Further, there is no net motion perpendicular to the force and thus $\chi_{J_2} = 0$.
The opposite behavior is observed for strong magnetic field $\alpha \gg 1$, where the Lorentz force causes the particle to move almost perpendicular to the applied force.
The observable $J_3$ saturates the bound at arbitrary magnetic field, i.~e.~it has the largest possible precision.
We note that for all observables and parameters, the bound $\chi_J \leq 1$ is always satisfied.
On the other hand, the TUR may be violated in the presence of a magnetic field, and the ratios $\tilde{\chi}_J$ for observables that involve motion perpendicular to the applied force can exceed unity and eventually diverge in the limit of strong magnetic field.
In this case, the entropy production no longer provides a valid bound on the precision of these observables.

\subsection{Brownian gyrator} \label{sec-examples-gyrator}
\begin{figure}
\includegraphics[width=.47\textwidth]{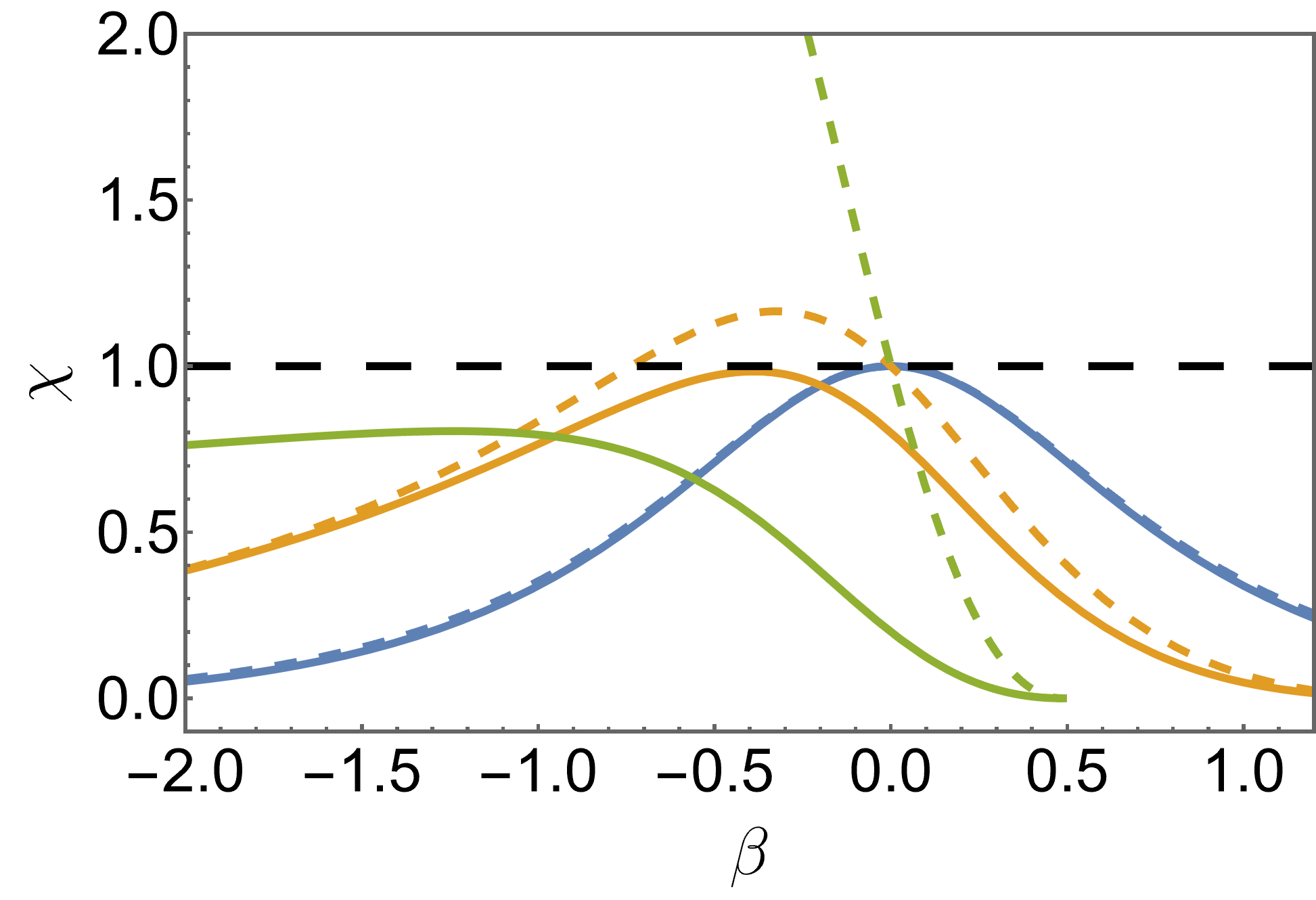}\\
\includegraphics[width=.47\textwidth]{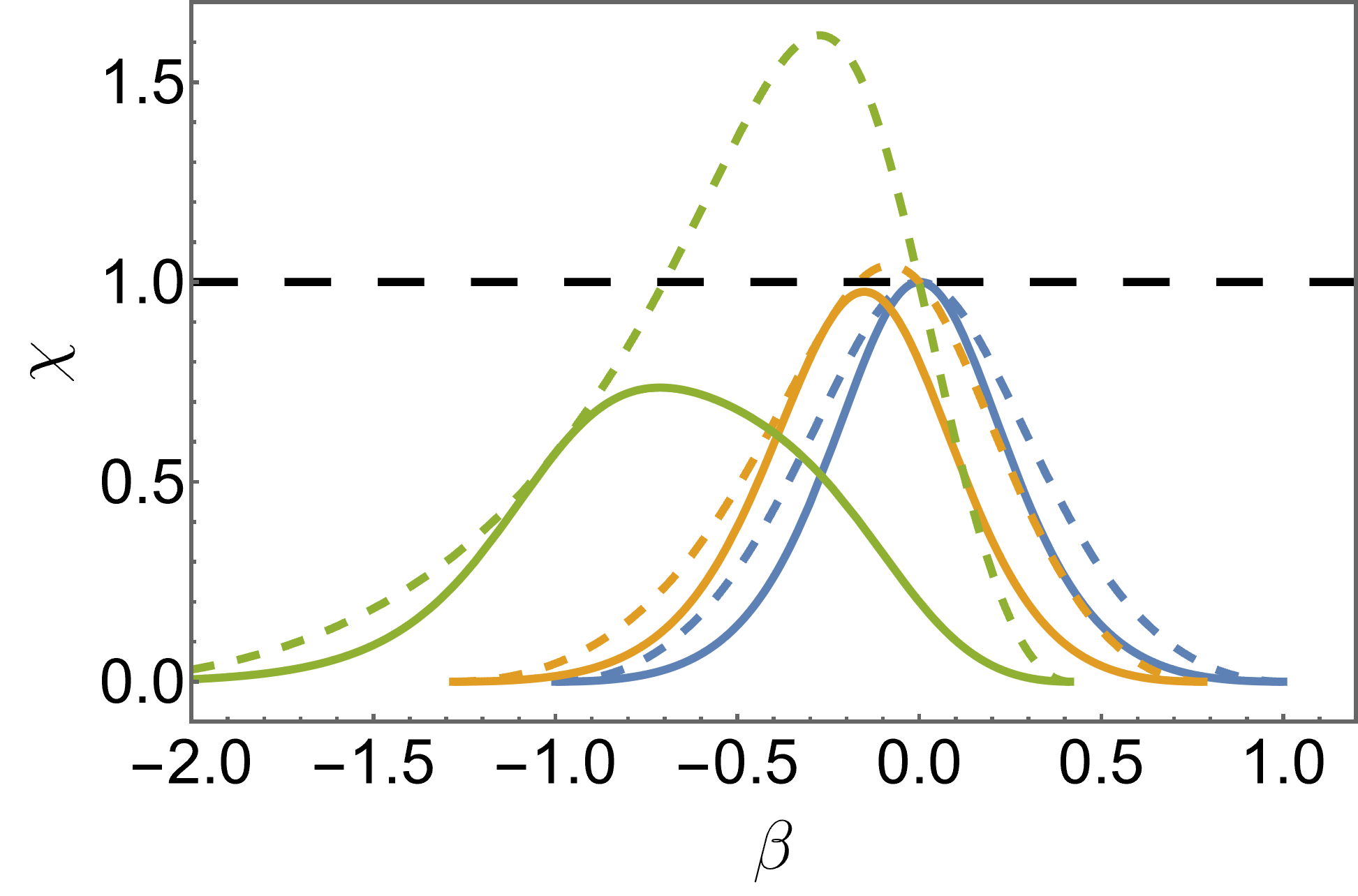}
\caption{The ratios $\chi_W$ (solid lines) and $\tilde{\chi}_W$ (dashed lines), \eqref{uncertainty-gyrator}, as a function of the driving strength $\beta$ for various values of the magnetic field. The blue lines are without magnetic field ($\alpha = 0$), while the orange ($\alpha = 0.5$) and green ($\alpha = 2$) ones correspond to weak and strong magnetic field, respectively. The top panel shows the results for small mass, $m_0 = 0.1$, the bottom panel for moderate mass $m_0 = 1$ \label{fig-gyrator}. Note that the lines terminate where the stability condition \eqref{gyrator-stability} is violated.}
\end{figure}
A more involved, yet still analytically solvable example is the Brownian gyrator \cite{Fil07}, which is a Brownian particle trapped in a harmonic potential and driven by a torque force.
This system has been studied also in the presence of a magnetic field \cite{Chu19} and it has been demonstrated that it can violate the thermodynamic uncertainty relation \eqref{TUR}.
For simplicity, we focus on the case where the magnetic field is along the $x_3$-direction, $\bm{B} = (0,0,B)$ and perpendicular to the torque force $\bm{F}^\text{t}(\bm{x}) = \kappa (-x_2,x_1,0)$.
We further take the confining harmonic potential to be isotropic, $\bm{F}^\text{c}(\bm{x}) = -k \bm{x}$.
In this case, the steady state probability density is \cite{Chu19}
\begin{align}
p(\bm{x},\bm{v}) = \bigg(\sqrt{\frac{m}{2 \pi T}}\bigg)^3 \exp\bigg[ - \frac{m\big\Vert \bm{v} - \bm{\nu}(\bm{x}) \big\Vert^2}{2 T} \bigg] p_x(\bm{x}),
\end{align}
with the marginal position density
\begin{align}
p_x(\bm{x}) &= \sqrt{\frac{k (k^*)^2 }{(2\pi T)^3}} \\
&\quad \times  \exp \Bigg[ -\frac{k \big(x_3 \big)^2 + k^* \big( x_1^2 + x_2^2 \big)}{2 T} \Bigg] \n 
\end{align}
and the local mean velocity
\begin{align}
\bm{\nu}(\bm{x}) = \frac{\kappa}{\gamma} \left(\begin{array}{c}
-x_2\\
x_1 \\
0 \end{array} \right) .
\end{align}
Here, in order to have a well-defined steady state, the system has to obey the stability criterion
\begin{align}
k^* = k - \kappa \bigg( \frac{q B}{\gamma} + \frac{m \kappa}{\gamma^2} \bigg) > 0 \label{gyrator-stability} ,
\end{align}
which intuitively states that the confining force has to be sufficiently strong compared to the driving torque force.
Once again, this system falls into the class of local equilibrium dynamics discussed in Sec.~\ref{sec-bounds-local}.
Evaluating the bound \eqref{generalized-TUR-local-magnetic} yields
\begin{align}
2 \eta_J &\leq 2\Bigg(t \bigg( \frac{\kappa^2}{\gamma T} + \frac{(\kappa q B)^2}{T \gamma^3} + \frac{4 m q B \kappa^3}{\gamma^4 T} + \frac{4 m^2 \kappa^4}{\gamma^5 T} \bigg) \nn
&\hspace{.5cm} + \frac{2 m \kappa^2}{\gamma^2 T} \Bigg) \frac{T}{k^*} .
\end{align}
We introduce the dimensionless parameters $\alpha = q B/\gamma$, $\beta = \kappa/k$ and $m_0 = k m/\gamma^2$ to write this as
\begin{align}
2 \eta_J &\leq\Big(t \big( 1 + (\alpha + 2 \beta m_0)^2 \big) + \frac{2 m }{\gamma} \Big) \sigma \label{TUR-gyrator} ,
\end{align}
where the entropy production rate $\sigma = \sigma^\text{od}$ is given by
\begin{align}
\sigma = \frac{2 \kappa^2}{\gamma k^*} .
\end{align}
$\alpha$ quantifies the strength of the magnetic field relative to the dissipation and $\beta$ the strength of the driving relative to the trapping.
$m_0$ can be written as $m_0 = \tau^\text{therm}/\tau^\text{relax}$, with $\tau^\text{therm} = m/\gamma$ the thermalization timescale of the velocity and $\tau^\text{relax} = \gamma/k$ the relaxation timescale in the confining potential. 
Thus, $m_0$ measures how fast the velocity thermalizes compared to the position dynamics, with small $m_0$ corresponding to the overdamped limit.
Compared to the overdamped TUR \eqref{TUR}, we see that the time-extensive term is multiplied by a factor $>1$, which recovers the TUR only for $\alpha = \beta m_0 = 0$, i.~e.~in the limit of no magnetic field and vanishing driving and/or mass.
This means that the bound \eqref{TUR-gyrator} leaves open the possibility for a violation of the TUR.
As we remarked above, this violation has explicitly been demonstrated in the presence of a magnetic field in \cite{Chu19}.
However, even if no magnetic field is involved ($\alpha = 0$) we have the bound
\begin{align}
2 \eta_J &\leq \Big(t \big( 1 + 4 \beta^2 m_0^2 \big) + \frac{2 m }{\gamma} \Big) \sigma \label{TUR-gyrator-no-magnetic},
\end{align}
whose right-hand side is larger than in the TUR.
On the other hand, the bound \eqref{bound-velocity} derived in Ref.~\cite{Van19} reads
\begin{align}
2 \eta_J &\leq \Big( 10 t + \frac{2 m }{\gamma} \Big) \sigma + \frac{\gamma t}{m} \label{TUR-gyrator-velocity} .
\end{align}
Comparing this to \eqref{TUR-gyrator-no-magnetic}, we see that the latter provides a tighter bound for small to moderate mass and driving, $\beta m_0 \leq 2/3$.
While \eqref{TUR-gyrator-no-magnetic} converges to the TUR for $m \rightarrow 0$, the right-hand side of \eqref{TUR-gyrator-velocity} diverges along with the dynamical activity.
As a concrete example for an observable, we consider the current
\begin{align}
J(t) = \int_0^t ds \ \bm{F}^\text{t}(\bm{x}(s)) \cdot \bm{v}(s) \equiv W(t) \label{work},
\end{align}
i.~e.~the work $W(t)$ performed by the torque force.
For this observable we have \cite{Chu19}
\begin{subequations}
\begin{align}
\av{W(t)} &= \frac{2 \beta^2 k T}{\gamma \big( 1 -  \beta \big( \alpha + \beta m_0 \big) \big)} t, \\
\text{Var}(W(t)) &\simeq \frac{2 \beta^2 k^2 T^2 \big(1 + \beta^2(1+3 m_0) - \alpha \beta^3 m_0 \big)}{\gamma \big( 1 -  \beta \big( \alpha + \beta m_0 \big) \big)^3} t ,
\end{align} \label{work-average-variance}%
\end{subequations}
where the expression for the variance is valid in the long-time limit.
Using this, we obtain the long-time ratios $\chi_W$ corresponding to \eqref{TUR-gyrator} and $\tilde{\chi}_W$ to \eqref{TUR},
\begin{subequations}
\begin{align}
\chi_W &\simeq \frac{\tilde{\chi}_W}{ 1 + (\alpha + 2 \beta m_0)^2}, \\
\tilde{\chi}_W &\simeq \frac{\big(1 - \beta (\alpha + \beta m_0) \big)^3}{1 + \beta^2 (1+3 m_0) - \alpha \beta^3 m_0 }.
\end{align} \label{uncertainty-gyrator}%
\end{subequations}
These quantities are shown for various combinations of $\alpha$, $\beta$ and $m_0$ in Fig.~\ref{fig-gyrator}.
We see that we have $\tilde{\chi}_W \leq 1$ in the absence of a magnetic field, $\alpha = 0$.
\begin{figure}
\includegraphics[width=.47\textwidth]{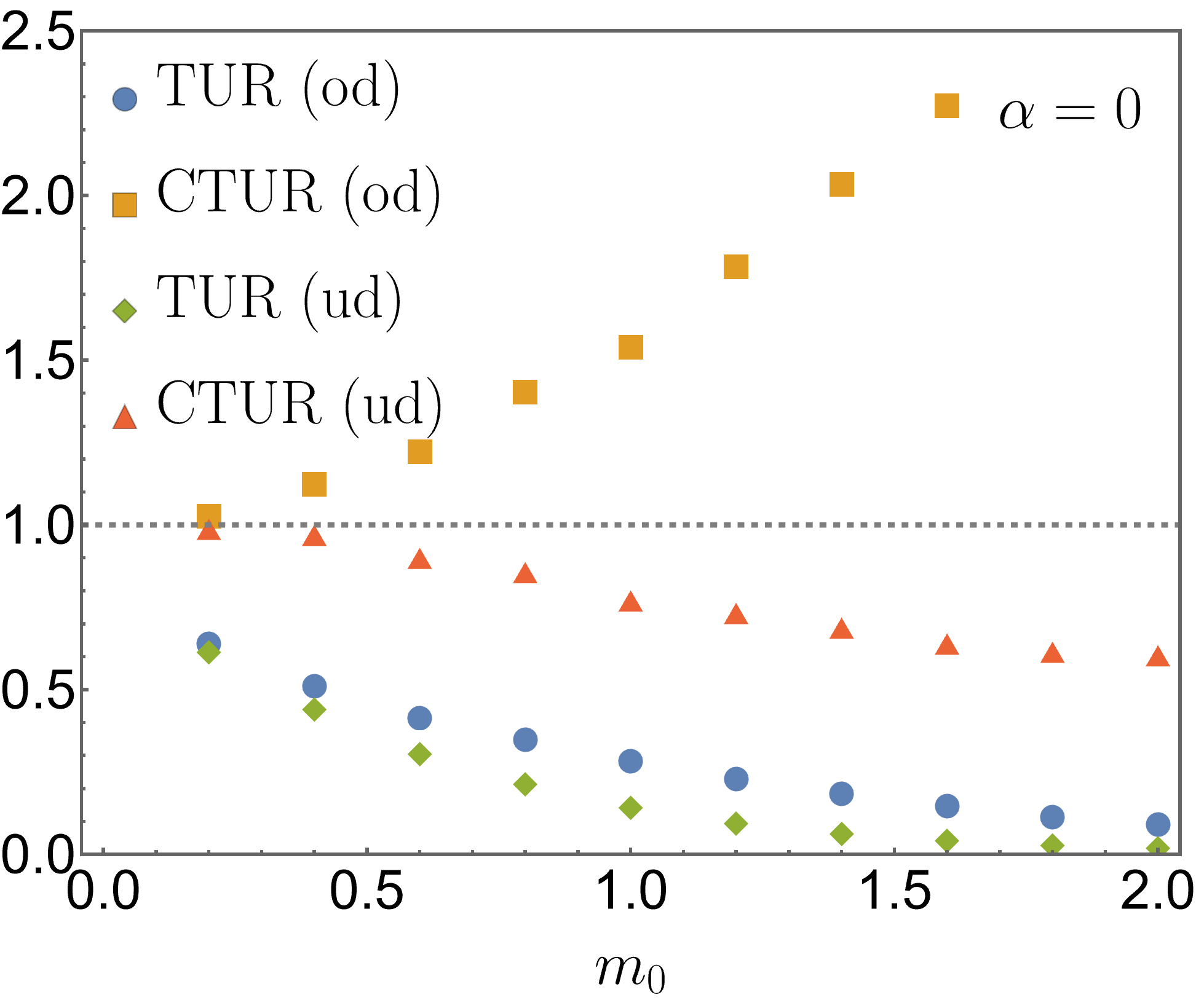}\\
\includegraphics[width=.47\textwidth]{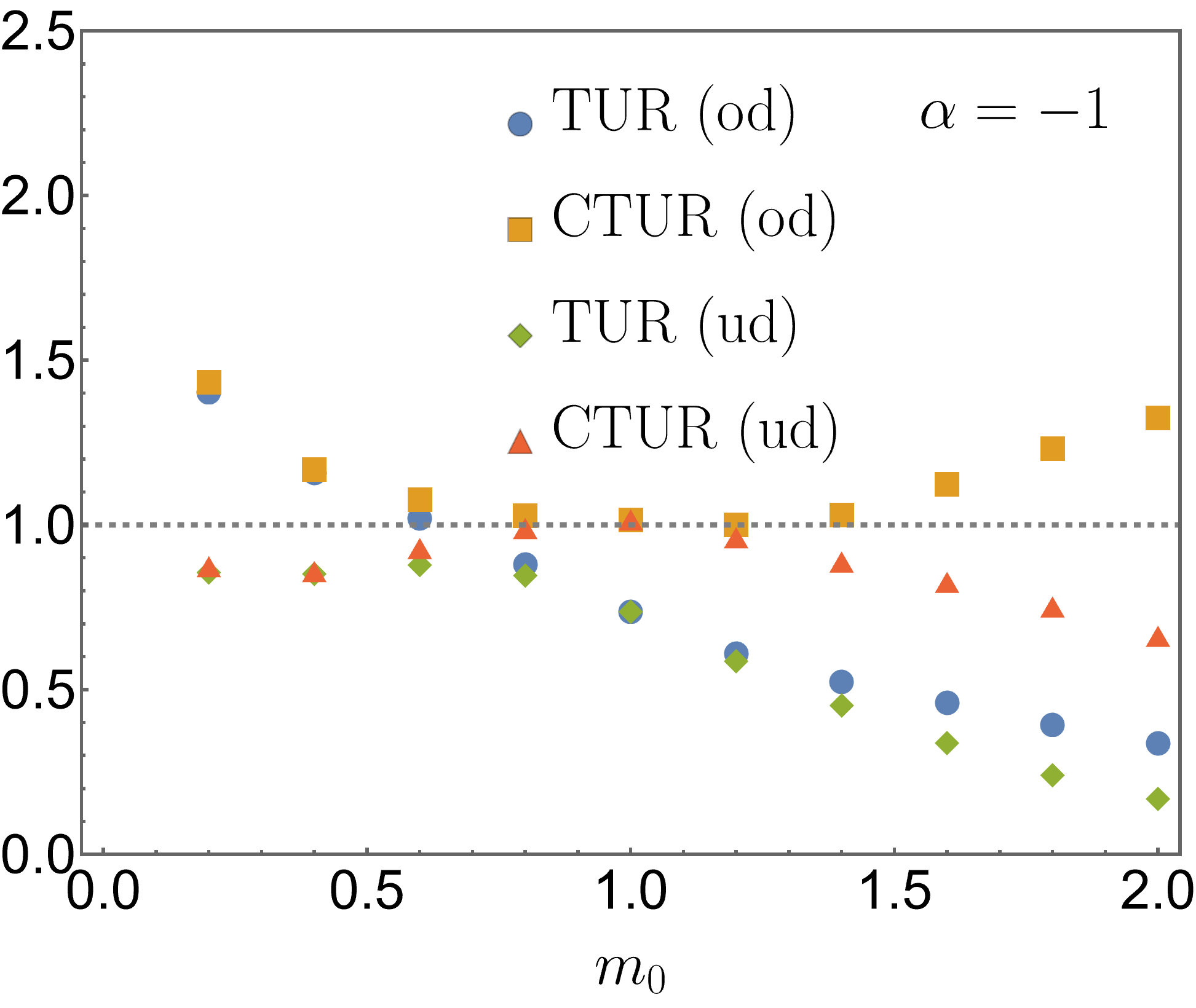}
\caption{Bounds on the precision of the work for the Brownian gyrator as a function of $m_0 = k m/\gamma^2$. The blue disks show the ratio between the left- and right-hand side of the overdamped TUR, $\tilde{\chi}_W = 2 \eta_w/\Delta S$, the orange squares show the same ratio for the CTUR, $2\eta_{W,\bar{W}}/\Delta S$. The green diamonds and red triangles are the respective ratios for the underdamped bounds, \eqref{main-bound-1} and \eqref{TUR-corr}. The parameters are $k = 2$, $\kappa = 1$, $\gamma = 1$, $T = 0.5$, and $q B = 0$ (top panel) and $q B = -1$ (bottom panel). The data are obtained by simulating $2 \cdot 10^4$ trajectories up to $t = 50$.\label{fig-gyrator-tur}}
\end{figure}
Thus, without a magnetic field, the work performed by the torque force satisfies the TUR even without the finite-mass terms in \eqref{TUR-gyrator}.
However, in the presence of a magnetic field, a violation of the TUR, $\tilde{\chi}_W > 1$, is observed when $\alpha$ and $\beta$ have opposite sign, i.~e.~when the Lorentz force pushes the particle towards the origin, counteracting the increase in the position variance due to the driving.
By contrast, for any set of parameter values, we always have $\chi_W \leq 1$, in accordance with the bound \eqref{TUR-gyrator}.
We remark that we cannot rule out that there may be other observables that violate the TUR even without a magnetic field, while still obeying the bound \eqref{TUR-gyrator-no-magnetic}.
In the case of the CTUR \eqref{TUR-corr}, we find that the overdamped bound can be violated with and without a magnetic field, see Fig.~\ref{fig-gyrator-tur}.
There, we perform numerical simulations of the Langevin equation describing the Brownian gyrator.
In addition to the mean and variance of the work, we also consider its correlations with the position-dependent observable
\begin{align}
\bar{W}(t) = \int_0^t ds \ \bm{F}^\text{t}(\bm{x}(s)) \cdot \bm{\nu}(\bm{x}(s)) ,
\end{align}
which characterizes the work done by a particle moving with the local mean velocity.
While in the overdamped regime, the left-hand side of \eqref{TUR-corr} provides an improved lower bound on the entropy production rate compared to the TUR \cite{Dec21}, in the underdamped case, this bound is no longer valid, even if no magnetic field is applied (top panel of Fig.~\ref{fig-gyrator-tur}).
However, the underdamped bound \eqref{TUR-corr} is always satisfied.
In the presence of a magnetic field, both the overdamped TUR and CTUR can be violated (bottom panel of Fig.~\ref{fig-gyrator-tur}), while the right-hand side of \eqref{TUR-gyrator} gives a valid bound in both cases.
We note that the underdamped bounds using the CTUR \eqref{TUR-corr} are relatively tight for all parameter values shown in Fig.~\ref{fig-gyrator-tur}.

We also have an alternative bound in terms of the equivalent equilibrium dynamics \eqref{generalized-TUR-eq-magnetic}, which for the present case reads
\begin{align}
\eta_J^\text{eq} &\leq\Big(t \big( 1 + \alpha^2 \big) + \frac{2 m }{\gamma} \Big) \sigma .
\end{align}
In the absence of a magnetic field, this implies the TUR \eqref{TUR} at long times,
\begin{align}
\eta_J^\text{eq} &\leq\Big(t + \frac{2 m }{\gamma} \Big) \sigma ,
\end{align}
with the caveat that the variance on the left-hand side has to be evaluated in the equivalent equilibrium system, corresponding to the conservative force field
\begin{align}
\bm{\phi}(\bm{x}) = - k \left(\begin{array}{c}
\Big( 1 - \frac{\kappa}{k} \big( \frac{q B}{\gamma} + \frac{m \kappa}{\gamma^2} \big) \Big) x_1\\
\Big( 1 - \frac{\kappa}{k} \big( \frac{q B}{\gamma} + \frac{m \kappa}{\gamma^2} \big) \Big) x_2 \\
 x_3 \end{array} \right),
\end{align}
which is just an anisotropic harmonic confinement.
For the observable \eqref{work}, the variance is easily evaluated,
\begin{align}
\text{Var}^\text{eq}(W(t)) &\simeq \frac{2 \beta^2 k^2 T^2}{\gamma \big( 1 -  \beta \big( \alpha + \beta m_0 \big) \big)} t.
\end{align}
Comparing this to \eqref{work-average-variance}, it can be checked that this is indeed smaller than the variance in the driven system and thus
\begin{align}
2 \eta_W &\leq 2 \eta_W^\text{eq} \leq\Big(t \big( 1 + \alpha^2 \big) + \frac{2 m }{\gamma} \Big) \sigma\label{TUR-gyrator-eq} .
\end{align}
We conjecture that the inequality $\text{Var}(J(t)) \geq \text{Var}^\text{eq}(J(t))$ also holds for more general observables, and thus that the TUR holds for a Brownian gyrator of arbitrary mass in the absence of a magnetic field.
Further note that the modification of the TUR in the presence of a magnetic field is the same as for driven diffusion, where the entropy production rate has to be multiplied by the factor $1+\alpha^2$ in order to obtain a valid bound.

\subsection{Diffusion in tilted periodic potentials} \label{sec-examples-periodic}
\begin{figure}
\includegraphics[width=.47\textwidth]{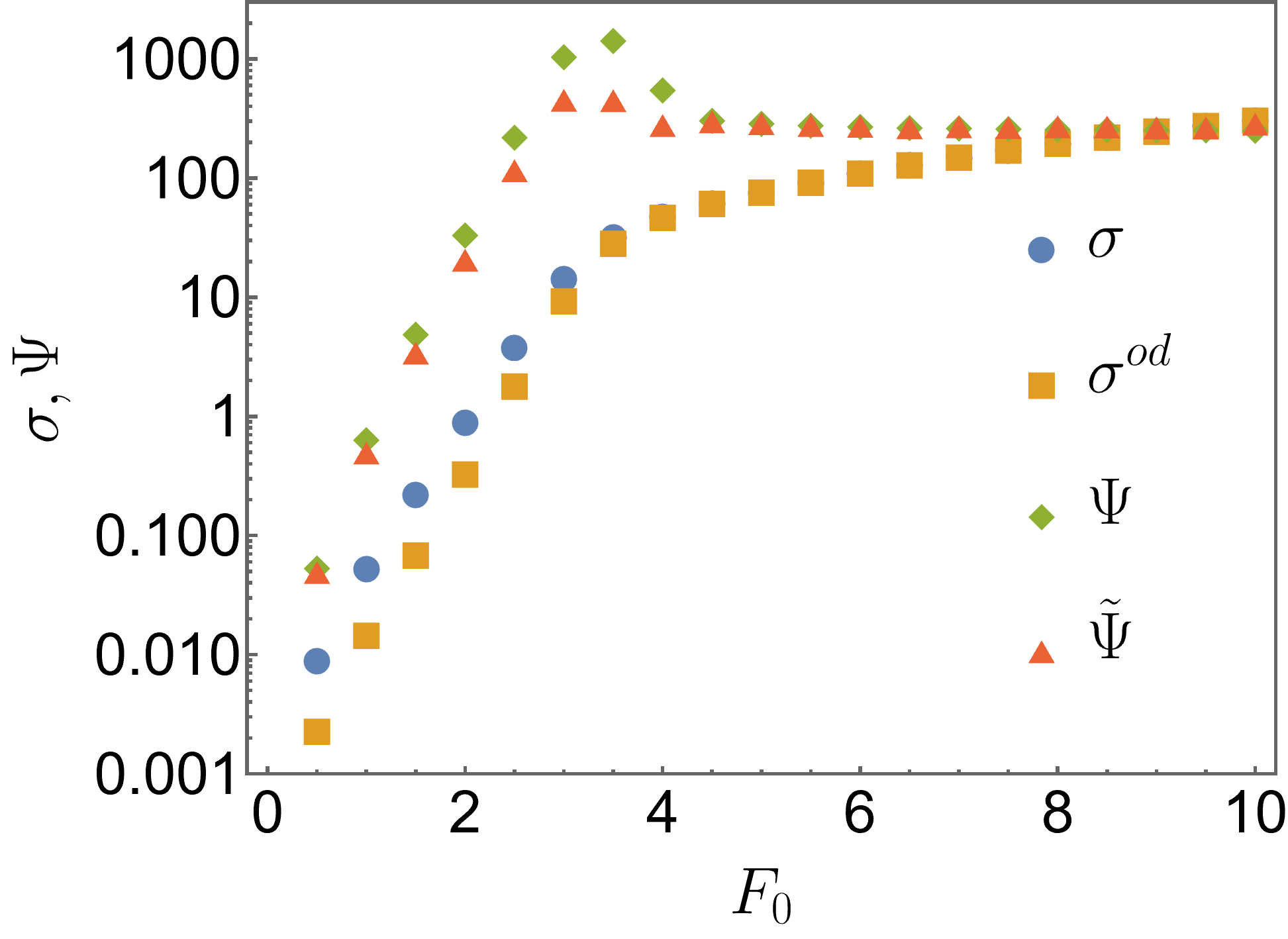}
\caption{The entropy production and the time-extensive correction term as a function of the bias $F_0$ for diffusion in a tilted periodic potential. The blue disks are the total entropy production \eqref{entropy-splitting}, the orange squares the overdamped contribution \eqref{entropy-over}. The green diamonds are the time-extensive correction term $\Psi$, computed from the actual velocity statistics according to \eqref{periodic-tur-velocity}, while the red triangles represent the result $\tilde{\Psi}$, \eqref{periodic-tur-local-mean}, obtained under the assumption of only the first cumulant depending on position. The parameters used in the simulation are $m = 0.5$, $\gamma = 1$, $T = 0.33$, $U_0 = 1$ and $L = 1$. We simulate $10^4$ trajectories up to time $t = 100$ after an initial equilibration phase of the same length. \label{fig-periodic-terms}}
\end{figure}
We now focus on the diffusion of a particle in a tilted, one-dimensional, periodic potential.
In this case the force in \eqref{langevin} is given by $F(x) = -U'(x) + F_0$, where $U(x) = U(x+L)$ is a periodic potential with lattice constant $L$ and $F_0$ is a constant bias force (tilt).
In the steady state of a one-dimensional system, we have from \eqref{kfp}
\begin{align}
\partial_x \big(\nu(x) p_x(x) \big) = 0 ,
\end{align}
i.~e., the product of local mean velocity and position density is constant.
Thus, we can write the local mean velocity as
\begin{align}
\nu(x) = \frac{\av{\nu}}{L p_x(x)} \label{meanvel-1d},
\end{align}
where $\av{\nu} = \av{v}$ is the steady-state average of the velocity.
This system was investigated in detail in Ref.~\cite{Fis20} and was conjectured to obey the TUR with an additional short time-correction.
Since an explicit expression for the probability density $p(x,v)$ is not available, we also cannot evaluate the quantity $h(x,v)$ determined by equation \eqref{h-equation} explicitly.
However, we may express it as a function of the conditional velocity density $p_v(v\vert x)$,
\begin{align}
h(x,v) = \partial_x \nu(x) v - \nu(x) \frac{\partial_x P_v(v \vert x)}{p_v(v \vert x)},
\end{align}
where $P_v(v \vert x)$ is the cumluative probability conditioned on $x$,
\begin{align}
P_v(v \vert x) = \int_{-\infty}^v du \ p_v(u \vert x) .
\end{align}
In the one-dimensional case, this can trivially be written as the velocity-gradient of another function, so it is also the optimal choice leading to the tightest bound.
Plugging this into \eqref{main-bound-1}, we obtain the bound
\begin{align}
2 \eta_J &\leq t \big(\sigma^\text{od} + \Psi\big) + \Omega_\nu \qquad  \text{with} \label{periodic-tur-velocity} \\ 
\Psi &= \frac{3 m^2 t}{\gamma T} \Av{\big(\nu \partial_x \nu \big)^2} \nn
&\qquad + \frac{m t}{\gamma} \Av{\Theta \big(\partial_x \nu \big)^2 + \nu \partial_x \nu \partial_x \Theta} \nn
& \qquad + \frac{m^2 t}{\gamma T} \Av{ \nu^2 \bigg(\frac{\partial_x P_v}{p_v} \bigg)^2} \nn
\Omega_\nu &= 2 \Av{\nu \partial_v \ln p_v \big)^2} \n .
\end{align}
Here, the function $\Theta(x)$, which we introduced in \eqref{theta}, quantifies the deviations of the local kinetic temperature from its equilibrium value,
\begin{align}
\Theta(x) = \frac{m}{T} \int dv \ \big(v - \nu(x)\big)^2 p_v(v \vert x) .
\end{align}
In the one-dimensional case, it is feasible to compute the joint position-velocity density $p(x,v)$ and the conditioned cumulative probability $P_v(v\vert x)$ directly from numerical simulations of the Langevin equation \eqref{langevin}.
By fitting the position-dependence of the latter with a periodic function, we can then compute its derivative and evaluate the right-hand side of the bound explicitly.
For comparison, we also evaluate the bound \eqref{main-bound-1-first-cumulant}, which was obtained under the assumption that $p_v(v \vert x)$ only depends on $x$ via its first cumulant $\nu(x)$.
Using \eqref{meanvel-1d}, this bound can be written as
\begin{align}
\tilde{\Psi} &= \frac{m t}{\gamma} \frac{\av{\nu}^2 \Theta}{L^2} \int_0^L dx \ \frac{\big(\partial_x p_x(x) \big)^2}{\big(p_x(x)\big)^3} \label{periodic-tur-local-mean} \\
&\qquad + \frac{4 m^2 t}{\gamma T} \frac{\av{\nu}^4}{L^4} \int_0^L dx \ \frac{\big(\partial_x p_x(x) \big)^2}{\big(p_x(x)\big)^5} \n ,
\end{align}
while the boundary term has the same form as in \eqref{periodic-tur-velocity}.
Here, the quantity $\Theta$ is defined as
\begin{align}
\Theta = \av{\Theta} = \frac{m}{T} \Av{(v-\nu)^2} .
\end{align}
The advantage of this expression is that it only requires the position density $p_x(x)$ and the average velocity fluctuations to be computed from the numerical simulations, however, the assumption of only the first cumulant of the velocity statistics depending on $x$ may not be justified.
Both results for the time-extensive term are shown in Fig.~\ref{fig-periodic-terms} as a function of the bias.
We use a simple sine-potential of the form
\begin{align}
U(x) = U_0 \sin\bigg(\frac{2 \pi x}{L} \bigg).
\end{align}
We observe that both for small and for large bias, \eqref{periodic-tur-velocity} and \eqref{periodic-tur-local-mean} yield the same value for the correction term.
This implies that in these regimes, the velocity statistics indeed depend on the position mainly via the local mean velocity $\nu(x)$.
Note that this does not imply local equilibrium statistics of the type \eqref{local-equilibrium}, for example for $F_0 = 3$, we find $\Theta \approx 3.5$ and, thus, velocity fluctuations well in excess of thermal equilibrium  fluctuations.
This is also reflected in the entropy production rate, where the second term in \eqref{entropy-splitting} contributes significantly for small bias, $\sigma \gg \sigma^\text{od}$.
In the intermediate regime, the correct expression for $\Psi$, \eqref{periodic-tur-velocity}, is significantly larger than \eqref{periodic-tur-local-mean}. 
This difference is due to the position dependence of the higher-order cumulants.

\begin{figure}
\includegraphics[width=.47\textwidth]{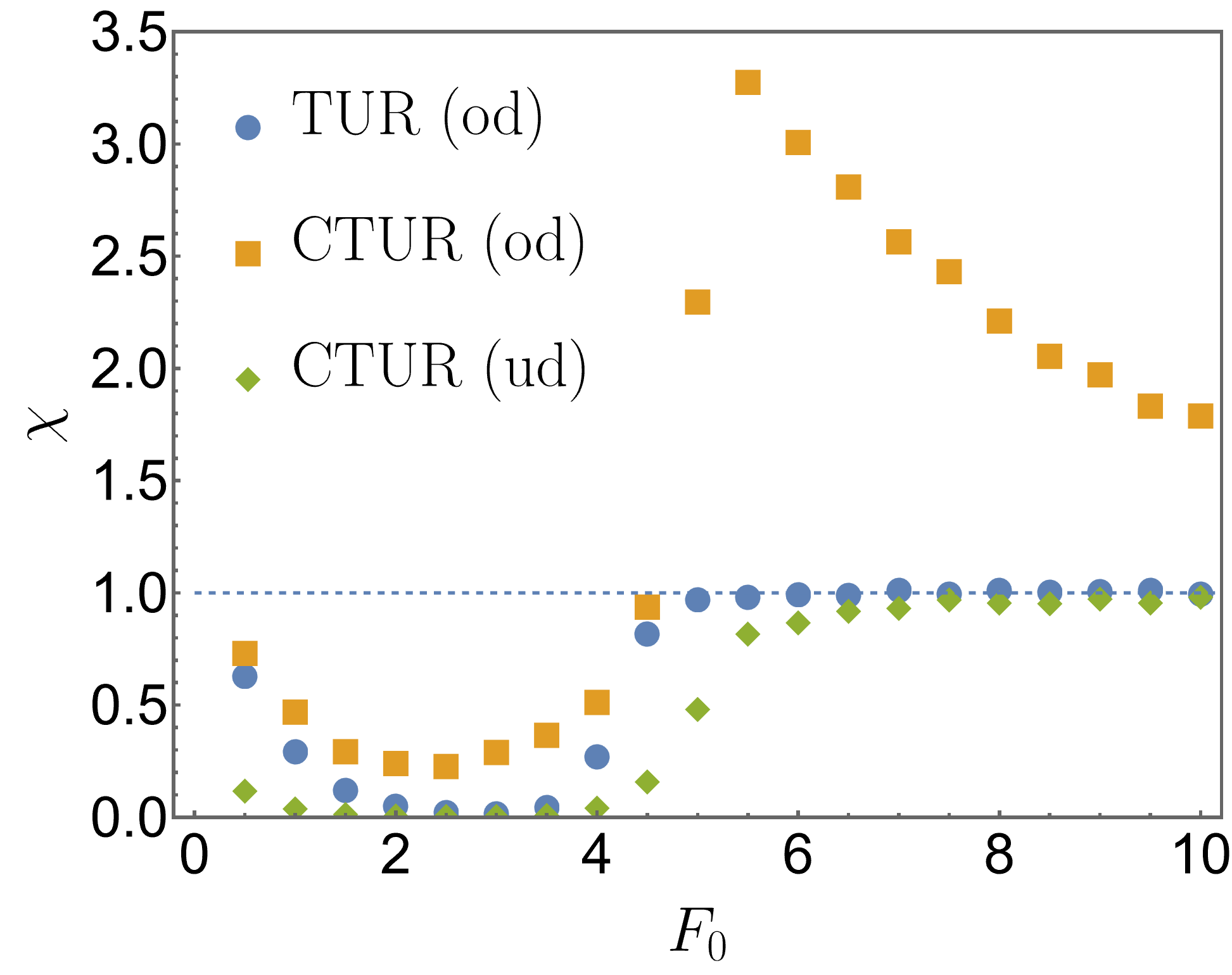}
\caption{The bound on the precision $\eta_J$ of the displacement as a function of the bias $F_0$ for diffusion in a tilted periodic potential. The blue disks correspond to the overdamped TUR \eqref{TUR}, $\tilde{\chi}_J = 2 \eta_J /\Delta S$. The orange squares correspond to the overdamped CTUR $\tilde{\chi}_{J,Z} = 2\eta_{J,Z}/\Delta S$, \eqref{TUR-corr}, while the green diamonds are the CTUR $\chi_{J,Z} = 2\eta_{J,Z}/\Sigma$ including the time-extensive correction term $\Sigma = t (\sigma^\text{od} + \Psi)$, \eqref{periodic-tur-velocity}. The parameters are the same as in Fig.~\ref{fig-periodic-terms}. \label{fig-periodic-tur}}
\end{figure}
In Fig.~\ref{fig-periodic-tur}, we show the bound on the precision of the displacement,
\begin{align}
J(t) = \int_0^t ds \ v(s) .
\end{align}
As was conjectured in Ref.~\cite{Fis20} based on numerical observations, the overdamped TUR \eqref{TUR} holds for this system in the long-time limit (blue disks).
Given that, one may wonder whether the additional terms in \eqref{periodic-tur-velocity} are necessary.
As we showed in Sec.~\ref{sec-multi-tur}, we can obtain a bound that is tighter than the TUR by additionally considering a position-dependent observable of the type \eqref{position-observable}.
Since, in the underdamped case, we do not have an explicit expression for the local mean velocity to compute \eqref{local-mean-current}, we use the observable
\begin{align}
Z(t) = \int_0^t ds \ \sin\bigg(\frac{2 \pi x(s)}{L} \bigg) .
\end{align}
For small and moderate bias, we observe the same behavior as for the overdamped case \cite{Dec21} (orange squares).
The information about correlations between the displacement $J(t)$ and the observable $Z(t)$ yields a improved bound on the entropy production compared to the TUR.
However, for larger bias, we see that the ratio $\eta_{J,Z}$ defined in \eqref{TUR-corr} is actually not a lower bound on the entropy production, as we observe $2 \eta_{J,Z} \geq \Delta S$.
On the other hand, the CTUR with the additional time-extensive term $\Psi$ given by \eqref{periodic-tur-velocity} remains valid and approaches unity for large bias.
This implies that the correction terms in \eqref{periodic-tur-velocity} are indeed necessary and that the bound can be tight.

\begin{figure}
\includegraphics[width=.47\textwidth]{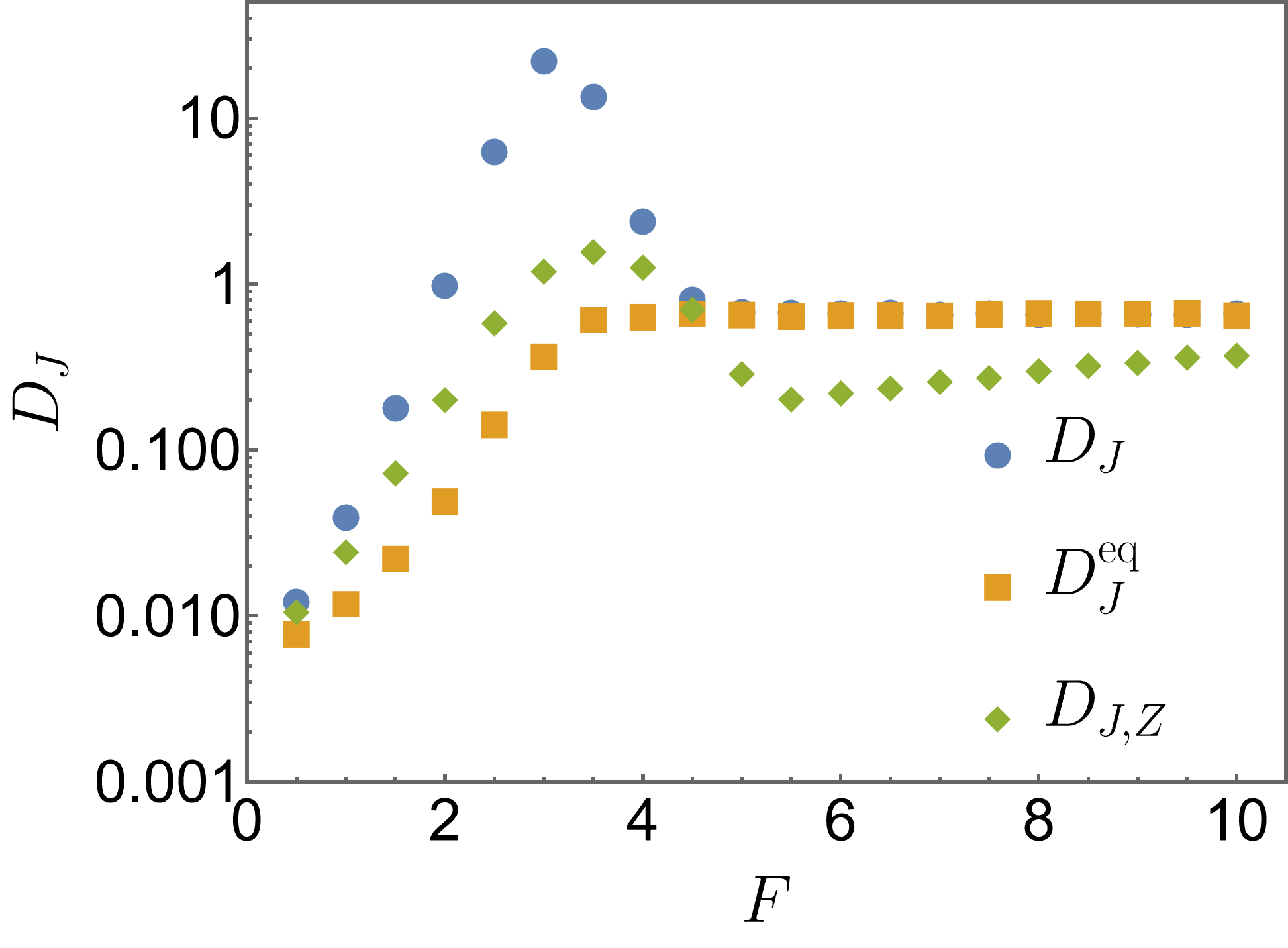}
\caption{The long-time diffusion coefficient $D_J = \lim_{t \rightarrow \infty} \text{Var}(J(t))/(2 t)$ as a function of the bias $F_0$ for diffusion in a tilted periodic potential. The blue disks are the diffusion coefficient of the displacement, evaluated in the dynamics \eqref{langevin}, while the orange squares are the diffusion coefficient for the equilibrium dynamics with the same position density \eqref{langevin-eq}. For comparison, we also show the modified diffusion coefficient $D_{J,Z} = \lim_{t \rightarrow \infty} (\text{Var}(J(t)) - \text{Cov}(J(t),Z(t))^2/\text{Var}(Z(t))/(2 t)$ (green diamonds), that is, the long-time limit of the expression in the denominator in the CTUR \eqref{TUR-corr}. The parameters are the same as in Fig.~\ref{fig-periodic-terms}. \label{fig-periodic-diff}}
\end{figure}
Next, we investigate the equilibrium fluctuation bound \eqref{generalized-TUR-eq}.
To do so, we fit the numerically obtained position probability density and calculate its logarithmic derivative \eqref{equilibrium-force}.
We then use the obtained force to perform simulations of \eqref{langevin-eq}.
Doing so, we find $\text{Var}^\text{eq}(J(t)) \leq \text{Var}(J(t))$, as is shown in Fig.~\ref{fig-periodic-diff}, so the equilibrium dynamics \eqref{langevin-eq} indeed exhibits reduced fluctuations in the current.
This implies that the equilibrium bound \eqref{generalized-TUR-eq} also bounds the precision of the current in the driven system.
However, since this system already satisfies the overdamped TUR, the bound is redundant in this case.
On the other hand, the left-hand side of the CTUR \eqref{TUR-corr} can be larger than the equilibrium bound, since we can have $\text{Var}(J(t)) - \text{Cov}(J(t),Z(t))^2/\text{Var}(Z(t)) \leq \text{Var}^\text{eq}(J(t))$.
The reason is that, in order to generalize the CTUR to equilibrium fluctuations, both the variance of the current and the correlations with the observable $Z(t)$ have to be evaluated in the equilibrium dynamics \eqref{langevin-eq},
\begin{align}
\frac{\av{J(t)}^2}{\text{Var}^\text{eq}(J(t)) - \frac{\text{Cov}^\text{eq}(J(t),Z(t))^2}{\text{Var}^\text{eq}(Z(t))}} \leq \frac{1}{2} \Sigma^\text{eq} \label{TUR-corr-eq}.
\end{align}
For the present case, we find that the current $J(t)$ and the observable $Z(t)$ are uncorrelated in the equilibrium dynamics, $\text{Cov}^\text{eq}(J(t),Z(t)) = 0$, so that \eqref{TUR-corr-eq} does not yield any improvement over the equilibrium bound \eqref{generalized-TUR-eq}.
This is in contrast to the driven system, where \eqref{TUR-corr} can be significantly tighter than \eqref{main-bound-1}.

\begin{figure}
\includegraphics[width=.47\textwidth]{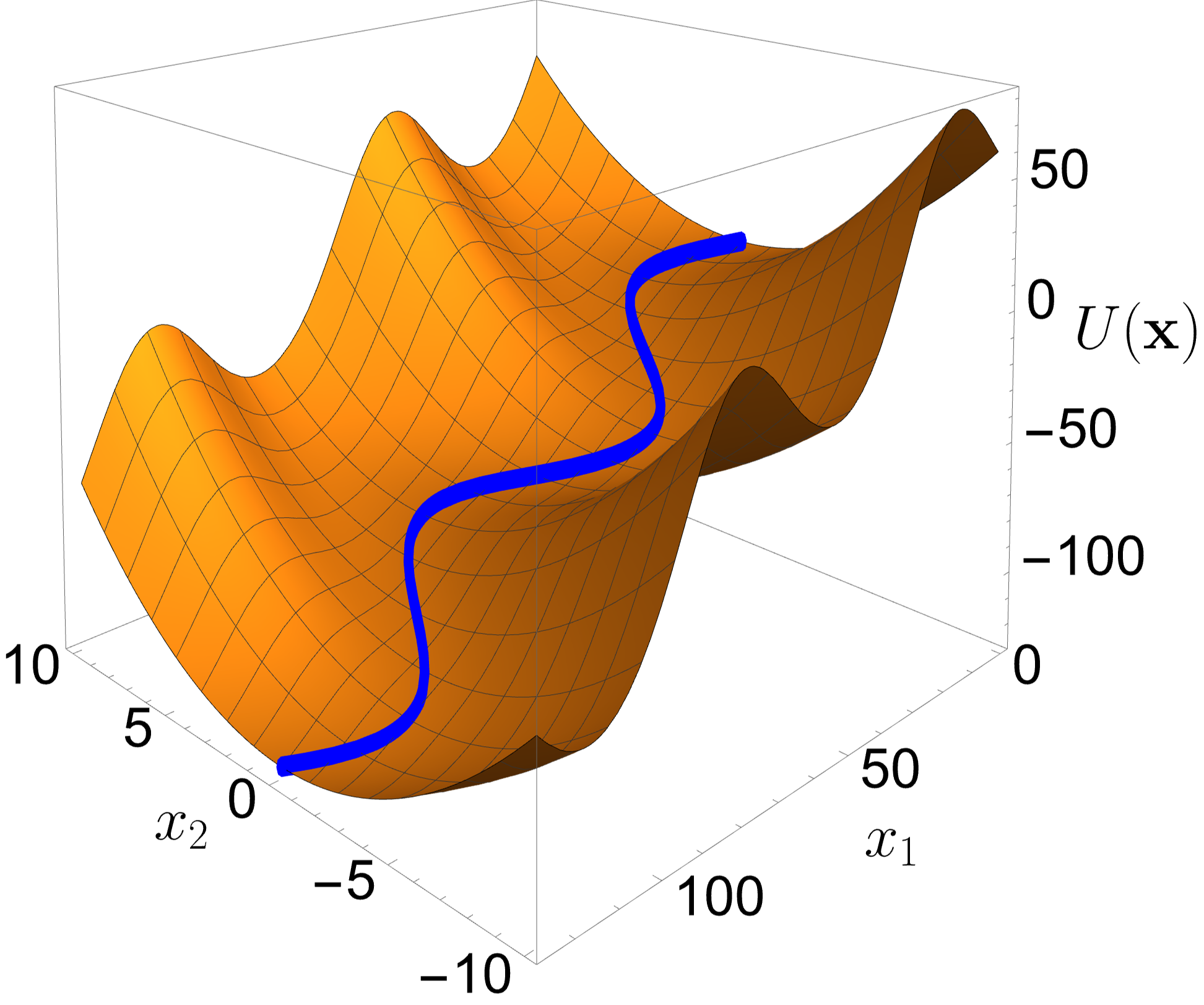}
\caption{The two-dimensional periodic potential \eqref{perpot-2D}. The blue line indicates the minimum of the potential in $x_2$-direction. The parameters are $k = 1$, $\kappa = 0.2$, $a = 3$, $L = 20 \pi$ and $F_0 = 1$. \label{fig-periodic2D-potential}}
\end{figure}
\begin{figure}
\includegraphics[width=.47\textwidth]{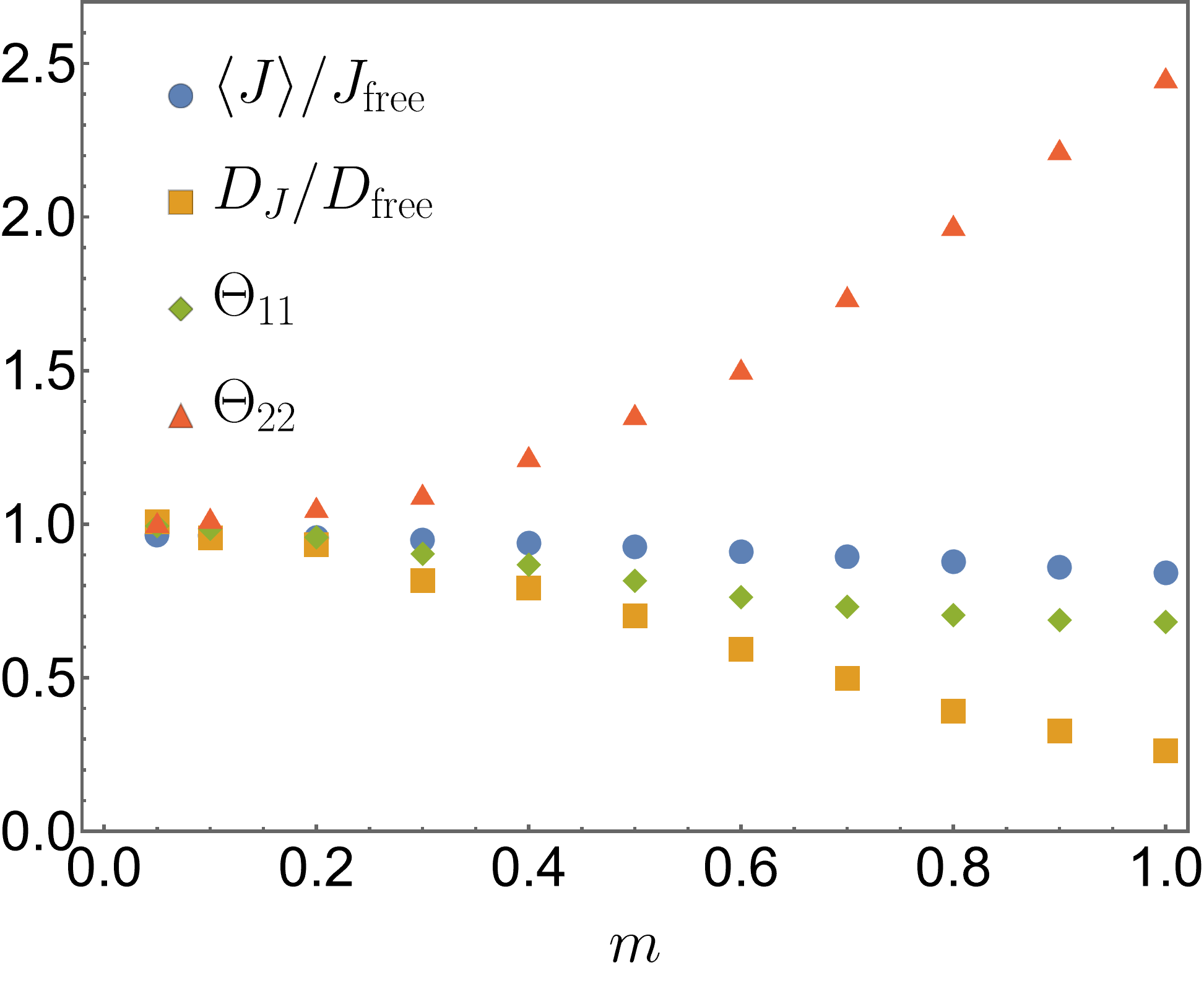}
\caption{The average displacement (blue disks) and diffusion coefficient (orange circles) relative to the free space values $J_\text{free} = F_0/\gamma$ and $D_\text{free} = 2 T/\gamma$ as a function of the particle mass $m$ for diffusion in the two-dimensional tilted periodic potential \eqref{perpot-2D}. Also shown are the velocity fluctuations relative to thermal fluctuations in the longitudinal (green diamonds) and transverse direction (red triangles). The parameters are $T = 1$, $\gamma = 0.1$, $k = 1$, $\kappa = 0.2$, $a = 3$, $L = 20 \pi$ and $F_0 = 1$.  \label{fig-periodic2D-diff}}
\end{figure}
Finally, while no counter-examples to the TUR \eqref{TUR} are known in the one-dimensional case \cite{Fis20}, it was recently shown in Ref.~\cite{Pie21} that in two- and higher-dimensional systems the TUR can be violated in a suitably designed potential landscape.
The mechanism for violating the TUR rests on exploiting the coherent oscillations of an underdamped particle to reduce the fluctuations of the measured current.
As a concrete example, we consider the two-dimensional potential discussed in Ref.~\cite{Pie21},
\begin{align}
U(\bm{x}) = \frac{k}{2} x_2^2 + \frac{\kappa}{2} \bigg( x_2 - a \sin \bigg(\frac{2 \pi x_1}{L} \bigg) \bigg)^2 \label{perpot-2D}.
\end{align}
This potential, which is shown in Fig.~\ref{fig-periodic2D-potential}, is periodic in the longitudinal $x_1$-direction and harmonic in the transverse $x_2$-direction.
In addition, we apply a constant bias $F_0$ in the longitudinal direction, which corresponds to tilting the potential in this direction.
The crucial feature of this potential is that the oscillations in the transverse direction constrain the motion in the longitudinal direction.
As discussed in Ref.~\cite{Pie21}, a particle, whose motion in the longitudinal direction is too fast compared to the transverse oscillations will find itself climbing the potential and thus be slowed down.
Similarly, if the particle is too slow, it will be accelerated by sliding down the potential.
In this way, the transverse oscillations focus the longitudinal motion, reducing the dispersion of the displacement without substantially reducing its average value, see Fig.~\ref{fig-periodic2D-diff}.
This phenomenon also manifests itself in the velocity fluctuations, which are suppressed in the longitudinal direction and enhanced in the transverse direction.
We remark that we can write the velocity-fluctuation part of the entropy production rate \eqref{entropy-splitting} in terms of the matrix $\bm{\Theta}(\bm{x})$ in \eqref{theta} as
\begin{align}
\sigma - \sigma^\text{od} = \frac{\gamma}{m} \big( \Av{\text{tr}(\bm{\Theta})} - N \big).
\end{align}
The positivity of this term implies that the suppression in one direction always has to be at least compensated by the enhancement in another direction; in Fig.~\ref{fig-periodic2D-diff} the sum of $\Theta_{11}$ and $\Theta_{22}$ is always larger than $2$.
As a consequence of the reduced fluctuations, the precision of the current can exceed the overdamped TUR \eqref{TUR}.
\begin{figure}
\includegraphics[width=.47\textwidth]{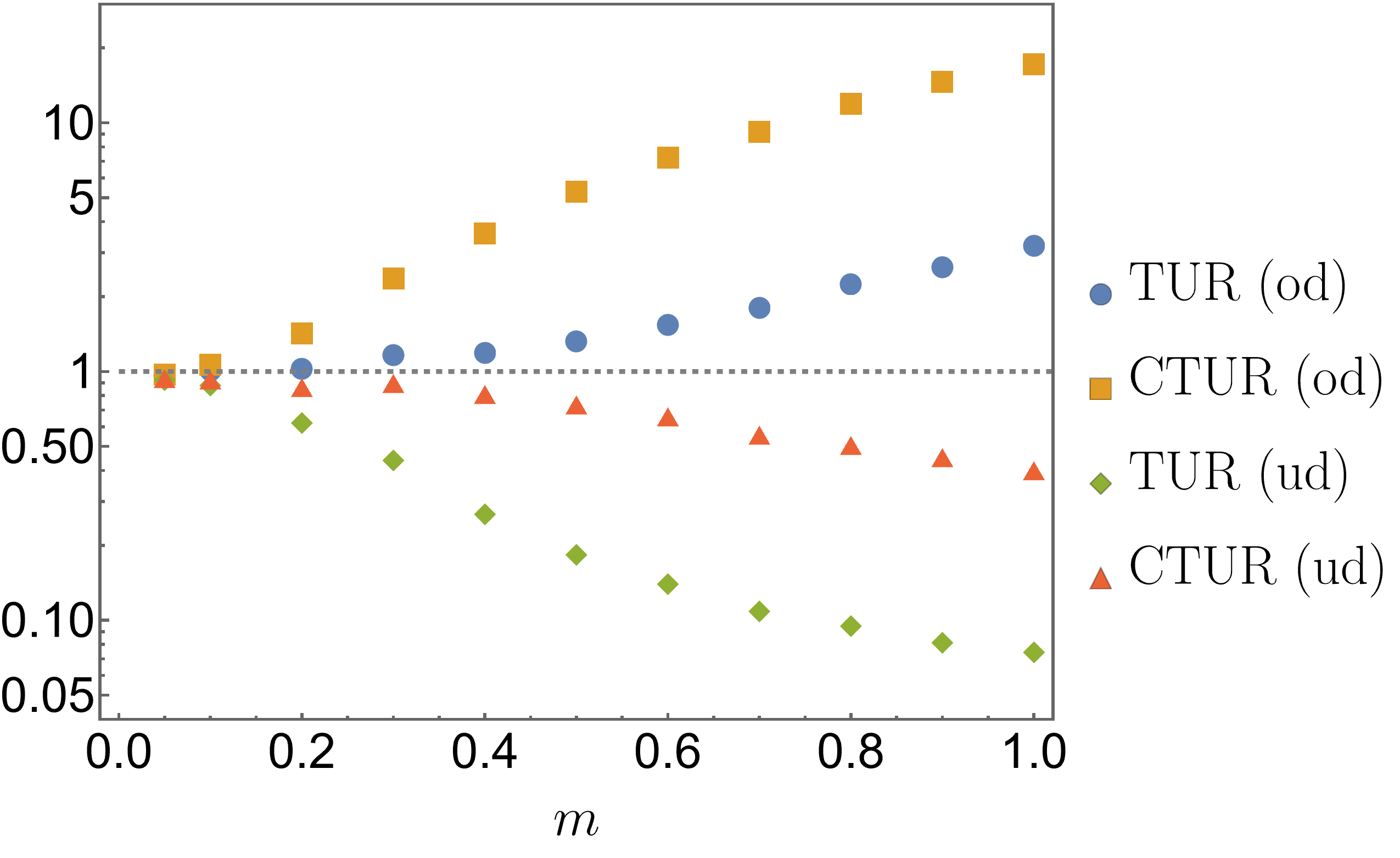}
\caption{Various bounds on the precision of the displacement of an underdamped particle in the potential \eqref{perpot-2D} as a function of the particle mass $m$. 
The blue disks correspond to the overdamped TUR, $\tilde{\chi}_J = 2 \eta_J/\Delta S$, the orange circles to the overdamped CTUR, $\tilde{\chi}_{J,Z} = 2 \eta_{J,Z}/\Delta S$.
The bounds involving the additional finite-mass terms \eqref{main-bound-1} and \eqref{TUR-corr} are indicated by the green diamonds and red triangles, respectively.
The parameters are $T = 1$, $\gamma = 0.1$, $k = 1$, $\kappa = 0.2$, $a = 3$, $L = 20 \pi$ and $F_0 = 1$; we perform numerical simulations of $2 \cdot 10^4$ trajectories up to time $50 m/\gamma$.  \label{fig-periodic2D-tur}}
\end{figure}
This can be seen in Fig.~\ref{fig-periodic2D-tur}, where the ratio $\tilde{\chi}_J = 2 \eta_J/\Delta S$ (blue disks) exceeds unity as the mass increases, demonstrating that, indeed, the precision of the current can be considerably larger the bound given by the overdamped TUR.
This is even more pronounced for the CTUR (orange squares).
In order for \eqref{TUR-corr} to give an improvement over the bound on $J(t)$ alone, we need to find an observable of the type \eqref{position-observable} that is strongly correlated with the current.
Looking at Fig.~\ref{fig-periodic2D-potential}, it is intuitively clear that particles moving at a large velocity will tend to move along the outer edge of the \enquote{track} defined by the potential, while slower particles will stick closer to the inner edge.
Using this intuition, we define the observable $Z(t)$ as
\begin{align}
Z(t) = \int_0^t ds \ x_2(s) \sin\bigg(\frac{2 \pi x_1(s)}{L} \bigg) .
\end{align}
The function under the integral approximately mimics the behavior of the velocity: 
It is large for a particle at the outer edge, and small for a particle at the inner edge of the potential track.
From Fig.~\ref{fig-periodic2D-tur}, we see that this intuition works well, the resulting ratio $\tilde{\chi}_{J,Z} = 2 \eta_{J,Z}/\Delta S$ is significantly larger than $\tilde{\chi}_J$ and greatly exceeds the overdamped bound.
To evaluate the corresponding bounds including the finite-mass terms \eqref{main-bound-1} and \eqref{TUR-corr}, in principle, we need to solve \eqref{h-equation-5}.
However, this is a formidable task: The first obstacle is to obtain an expression for the conditional velocity density $p_v(\bm{v} \vert \bm{x})$.
This is challenging even using trajectory simulations, since the phase-space is four-dimensional and we need very good statistics to obtain this quantity with sufficient accuracy to compute its derivatives.
However, for the present example, we expect the bound \eqref{main-bound-1-first-cumulant}, where only the first cumulant of the velocity statistics depends on position, to be a good approximation.
There are two reasons for this: 
First, from the numerical simulations, we observe that, indeed, the position-dependence of the matrix $\bm{\Theta}(\bm{x})$ is relatively weak, and it is reasonable to assume that this also holds for the higher-order cumulants.
Second, for the choice of parameters used in Fig.~\ref{fig-periodic2D-tur}, the current and thus the magnitude of the local mean velocity are relatively large.
Since in \eqref{main-bound-1}, the term involving the local mean acceleration is proportional to the fourth power of the local mean velocity, while all other terms are generally of second order (see also \eqref{generalized-TUR-second-cumulant}), we expect this term to be dominant compared to the one characterizing the velocity fluctuations.
This fact is corroborated by the numerical simulations, where the local mean acceleration term is about two orders of magnitude larger than the velocity-fluctuation term.
The bound \eqref{main-bound-1-first-cumulant} can be evaluated without knowing the precise form of the velocity statistics by computing the local mean velocity $\bm{\nu}(\bm{x})$ and the matrix $\bm{\Theta}$ from the numerical simulations, which is a more feasible task.
The results are shown in Fig.~\ref{fig-periodic2D-tur}.
We see that indeed, the bound \eqref{main-bound-1-first-cumulant} holds for all values of the particle mass.
For small mass, both the overdamped and underdamped bounds are saturated, which implies that, in this limit, the motion in the longitudinal direction is well described by free diffusion.
For larger mass, \eqref{main-bound-1} is not tight, while the CTUR \eqref{TUR-corr} remains relatively tight.

\section{Discussion}
As the examples discussed in the previous sections show, the bound \eqref{main-bound-1} can be tight, provided that we take into account the correlations between the current and a suitably chosen position-dependent observable.
In the overdamped limit, as was discussed in Ref.~\cite{Dec21}, the CTUR \eqref{TUR-corr} always becomes an equality when choosing the stochastic entropy production and its local mean value as the observables, whereas the bound is generally tight for observables that approximate these quantities.
The reason for this tightness are the special properties of the overdamped stochastic entropy production \cite{Pig17,Dec21b}.
It is an intriguing question whether similar optimal observables exist in the underdamped case and what their implications on the stochastic entropy production are.

While in the overdamped case, it has been suggested to use the thermodynamic uncertainty relation as a way to estimate the entropy production \cite{Li19,Man20,Ots20,Van20}, the presence of the additional terms on the right-hand side of \eqref{main-bound-1} precludes such an approach in the underdamped regime.
However, the fact that the underdamped bounds can likewise be tight means that, in principle, we can estimate the right-hand side of \eqref{main-bound-1} using the measurement of stochastic currents and other observables.
In particular for systems where the magnitude of the local mean velocity is large, we expect the term involving the local mean acceleration to give the dominant contribution, since it is of fourth order in the local mean velocity.
In these cases, rather than yielding a lower bound on the entropy production rate, or, equivalently, the magnitude of the local mean velocity, we obtain a lower bound on the magnitude of the local mean acceleration, which may be interesting in its own right.

Finally, we note that since the additional terms on the right-hand side of \eqref{main-bound-1} involve spatial derivatives of the local mean velocity and the velocity probability density, a necessary property of underdamped systems violating the TUR is a pronounced spatial structure of the system.
This agrees with the arguments made in Ref.~\cite{Pie21}, where the coupling between different degrees of freedom---here, the motion along different spatial directions---was identified as the main ingredient for constructing counter-examples to the TUR.
The bound \eqref{main-bound-1} casts this intuitive reasoning into a quantitative form, which may be helpful in designing systems in such a way as to maximize their precision.

\begin{acknowledgments}
This work was supported by JSPS KAKENHI, grant number 19H05795. The author gratefully acknowledges discussions with Shin-ichi Sasa on the validity of the TUR in the underdamped regime.
\end{acknowledgments}


%

\onecolumngrid

\appendix

\section{Optimal perturbation force} \label{app-optimization}
In \eqref{h-equation-5}, we used the condition that the perturbation force $\bm{f}(\bm{x},\bm{v})$ minimizing the right-hand side of \eqref{fri-bound} should be a velocity-gradient, $\bm{f}(\bm{x},\bm{v}) = \grad_v \psi_f(\bm{x},\bm{v})$.
Minimizing the expression on the right-hand side of \eqref{fri-bound} under the constraint that it satisfies \eqref{f-equation} corresponds to minimizing the functional
\begin{align}
\int d\bm{x} \int d\bm{v} \ \big\Vert \bm{f}(\bm{x},\bm{v}) \big\Vert^2 p(\bm{x},\bm{v}) + 2 \int d\bm{x} \int d\bm{v} \ \psi_f(\bm{x},\bm{v}) \Big( \grad_v \cdot \big( \bm{f}(\bm{x},\bm{v}) p(\bm{x},\bm{v}) \big) - \chi(\bm{x},\bm{v}) \Big) ,
\end{align}
where $\chi(\bm{x},\bm{v})$ is the right-hand side of \eqref{f-equation}, which is independent of $\bm{f}(\bm{x},\bm{v})$, and $\psi_f(\bm{x},\bm{v})$ is a Lagrange multiplier.
Minimizing this with respect to $\bm{f}(\bm{x},\bm{v})$ yields the Euler-Lagrange equations
\begin{align}
2 \bm{f}(\bm{x},\bm{v}) p(\bm{x},\bm{x}) - p(\bm{x},\bm{v}) \grad_v \psi(\bm{x},\bm{v}) = 0. 
\end{align}
Since $p(\bm{x},\bm{v})$ is positive, this is equivalent the gradient condition $\bm{f}(\bm{x},\bm{v}) = \grad_v \psi_f(\bm{x},\bm{v})$.

\section{Perturbation force with a magnetic field} \label{app-magnetic-force}
Similar to \eqref{kfp-first-order}, the leading order contribution $\pi(\bm{x},\bm{v})$ to the change in the probability density $p(\bm{x},\bm{v})$ in the presence of a magnetic field is given by
\begin{align}
\mathcal{L}(\bm{x},\bm{v}) \pi(\bm{x},\bm{v}) &= \frac{1}{m} \grad_v \big( \bm{f}(\bm{x},\bm{v}) p(\bm{x},\bm{v})\big) \qquad \text{with} \label{force-condition-magnetic} \\
\mathcal{L}(\bm{x},\bm{v}) &= -\bm{v} \cdot \grad_x - \frac{1}{m} \grad_v \cdot \Big( \bm{F}(\bm{x}) + q \bm{v} \times \bm{B}(\bm{x}) - \gamma \bm{v} - \frac{\gamma T}{m} \grad_v \Big) \n ,
\end{align}
where $\bm{f}(\bm{x},\bm{v})$ is the small perturbation of the force.
As before, we want to determine the force $\bm{f}(\bm{x},\bm{v})$ such that
\begin{align}
\pi(\bm{x},\bm{v}) = -\bm{\nu}(\bm{x}) \cdot \grad_v p(\bm{x},\bm{v}),
\end{align}
see \eqref{probability-mod}.
Since we know that $p(\bm{x},\bm{v})$ is the steady-state solution of \eqref{kfp-magnetic},
\begin{align}
\mathcal{L}(\bm{x},\bm{v}) p(\bm{x},\bm{v}) = 0,
\end{align}
we can use this to simplify the left-hand side of \eqref{force-condition-magnetic},
\begin{align}
\mathcal{L}(\bm{x},\bm{v}) \pi(\bm{x},\bm{v}) = \frac{1}{m} \grad_v \cdot \bigg(  \Big( \gamma \bm{\nu}(\bm{x}) - q \bm{\nu}(\bm{x}) \times \bm{B}(\bm{x}) \Big) p(\bm{x},\bm{v}) \bigg) + \grad_v \cdot \Big(p(\bm{x},\bm{v}) \big(\bm{v} \cdot \grad_x \big) \bm{\nu}(\bm{x}) \Big) - \grad_x \cdot \big(\bm{\nu}(\bm{x}) p(\bm{x},\bm{v}) \big) .
\end{align}
Writing the force as
\begin{align}
\bm{f}(\bm{x},\bm{v}) = \gamma \bm{\nu}(\bm{x}) - q \bm{\nu}(\bm{x}) \cdot \bm{B}(\bm{x}) + m \bm{h}(\bm{x},\bm{v}),
\end{align}
we obtain \eqref{h-equation-magnetic}.

\end{document}